\documentclass[twocolumn,prl,tightenlines,superscriptaddress]{revtex4-2}

\usepackage{hyperref}
\usepackage{amsmath}
\usepackage{amssymb,amsfonts,latexsym}
\usepackage{bm}
\usepackage[mathcal]{euscript}
\usepackage{graphicx}
\usepackage{epsfig}
\usepackage{color}
\usepackage{xfrac}
\usepackage{mathrsfs}

\begin{document}

\title{Integer Topological Defects Reveal Anti-Symmetric Forces in Active Nematics}

\author{Zihui Zhao*}
\affiliation{School of Physics and Astronomy, Institute of Natural Sciences, Shanghai Jiao Tong University, Shanghai 200240, China}
\altaffiliation[]{contributed equally}

\author{Yisong Yao*}
\affiliation{School of Physics and Astronomy, Institute of Natural Sciences, Shanghai Jiao Tong University, Shanghai 200240, China}
\altaffiliation[]{contributed equally}

\author{He Li}
\affiliation{School of Physics and Astronomy, Institute of Natural Sciences, Shanghai Jiao Tong University, Shanghai 200240, China}
\affiliation{Institut de Génétique et de Biologie Moléculaire et Cellulaire, Illkirch, France}

\author{Yongfeng Zhao}
\affiliation{School of Physics and Astronomy, Institute of Natural Sciences, Shanghai Jiao Tong University, Shanghai 200240, China}
\affiliation{Center for Soft Condensed Matter Physics \& Interdisciplinary Research, Soochow University, Suzhou 215006, China}

\author{Yujia~Wang}
\affiliation{Center for Soft Condensed Matter Physics \& Interdisciplinary Research, Soochow University, Suzhou 215006, China}

\author{Hepeng Zhang}
\affiliation{School of Physics and Astronomy, Institute of Natural Sciences, Shanghai Jiao Tong University, Shanghai 200240, China}

\author{Hugues Chat\'{e}}
\affiliation{Service de Physique de l'Etat Condens\'e, CEA, CNRS Universit\'e Paris-Saclay, CEA-Saclay, 91191 Gif-sur-Yvette, France}
\affiliation{Computational Science Research Center, Beijing 100094, China}

\author{Masaki Sano}
\affiliation{School of Physics and Astronomy, Institute of Natural Sciences, Shanghai Jiao Tong University, Shanghai 200240, China}
\affiliation{Universal Biology Institute, The University of Tokyo, Bunkyo-ku, Tokyo 113-0033, Japan}

\date{\today}

\begin{abstract}
Cell layers are often categorized as contractile or extensile active nematics 
but recent experiments on neural progenitor cells with induced $+1$ topological defects challenge this classification.
In a bottom-up approach, we first study a relevant particle-level model and then analyze a continuous theory derived from it. We show that both model and theory account qualitatively for the main experimental result, i.e. accumulation of cells at the core of any type of +1 defect. We argue that cell accumulation is essentially due to two generally ignored anti-symmetric active forces. We finally discuss the relevance and consequences of our findings in the context of other cellular active nematics experiments and previously proposed theories.
\end{abstract}

\maketitle

It is widely recognized today that topological defects can play a crucial role in biological contexts, including intracellular cytoskeletal dynamics, tissue growth during embryogenesis, and population-level expansion in bacterial colonies~\cite{Prost2004, Marchetti2013, Shankar2022, gompper20202020, Duclos2017, Copenhagen2021, Maroudas-Sacks2021}. 
Two-dimensional (2D) tissues formed by cells reaching confluence are a case of particular interest since they are often
precursor of 3D shape.
Tissues are often described as active nematics, usually because the elongated shapes of their cells give rise to local nematic order \cite{duclos2014,Duclos2018}. 
Such cellular active nematics, like their passive counterparts, can exhibit $\pm\tfrac{1}{2}$ half-integer topological defects, 
i.e. points in space around which the nematic director turns by $\pm\pi$.
Activity endows $+ \frac{1}{2}$ defects with an intrinsic velocity, and it is now well-known that 
cells can be attracted to their core, where they may nucleate
additional layers~\cite{Copenhagen2021,sarkar2023crisscross} or be extruded~\cite{hakim2017,Saw2017} or form 3D mounds~\cite{Kawaguchi2017}, 
while they are depleted at the core of $- \frac{1}{2}$ defects~\cite{Kawaguchi2017,Copenhagen2021}

Integer-charge topological defects can also play a role in biological development~\cite{Hoffmann2022,ho2024role,mietke2019self,mietke2019minimal} 
such as head and foot regeneration in hydra~\cite{Maroudas-Sacks2021} or the growth of plants around their meristems~\cite{hamant2008}.
They do not appear spontaneously in cellular active nematics, essentially for the same `energetic' reasons
as in equilibrium liquid crystals \cite{Schandrasekhar2013}. 
Charge $+1$ defects can, though, be induced either by confinement within small domains
~\cite{Guillamat, Blanch-Mercader2021PRL, Blanch-Mercader2021E},
or by gentle large-scale cues imprinted in the underlying substrate~\cite{Endresen2021,Kaiyrbekov2023,Zihuiexper}. 
In most cases reported so far, cell accumulation and/or extrusion in 3D has been observed at the core of $+1$ defects.

Models and theories of active nematics have been proposed~\cite{Marchetti2013,ramaswamy2003active,marenduzzo2007steady,giomi2011excitable,giomi2013defect,Shi2013,gao2015multiscale,giomi2015geometry,Green2017,Maitra2018,doostmohammadi2018}. 
Most adopt a `top-down' approach in terms of 
continuous fields ---usually a velocity or polarity field ${\bm v}$ and a nematic tensor field ${\bm Q}$---, 
and invoke a contractile/extensile dichotomy.  The situation is somewhat confusing: the cells involved
are frequently termed contractile.
Yet, at the collective level, 
cellular active nematics are often~\cite{Kawaguchi2017} 
---but not always~\cite{Duclos2017,balasubramaniam2021investigating}--- classified extensile 
from the direction of the intrinsic motion of $+\tfrac{1}{2}$ defects and the flow field around them. 
These comet-like structures can indeed be found to move `toward their head' like in standard theories
with an (extensile) active stress of negative coefficient $-\zeta{\bm Q}$ ($\zeta>0$) \cite{simha2002,Marchetti2013}. 
Progress toward understanding this conundrum was recently made by invoking the effect of `polar fluctuations'~\cite{Vafa2021,balasubramaniam2021investigating,Killeen2022, Zhang2023}.

\begin{figure}[t!]
\includegraphics[keepaspectratio,width=\columnwidth]{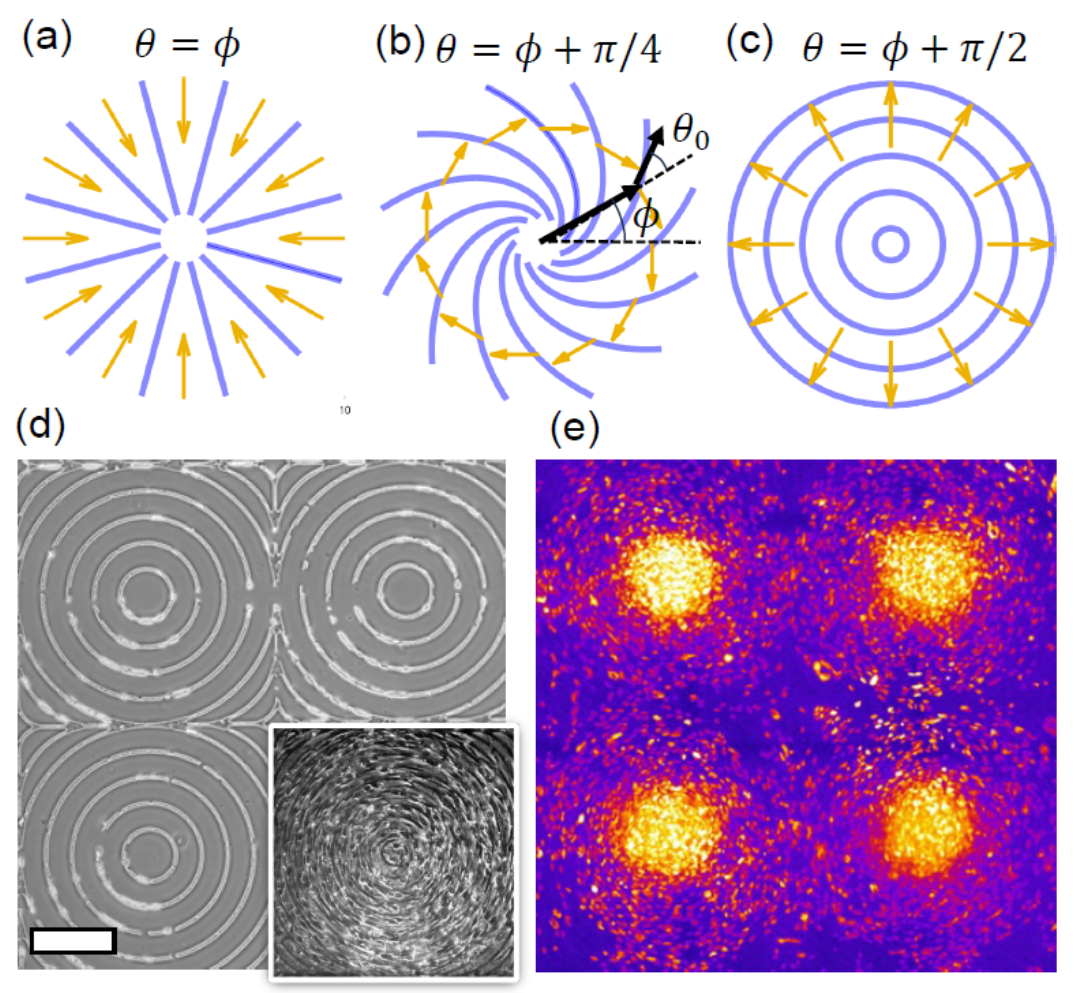} %
\caption{\label{fig1} 
(a-c): Sketch of $+1$ nematic defects and direction of the linear active force $-\zeta \nabla \cdot {\bm Q}$ when $\zeta>0$  (arrows); (a) aster, (b) spiral ($\theta_0=\pi/4$), (c) target. 
(d): Substrate with micro-fabricated shallow ridges (width $9\mu m$ and height $1.5\mu m$) that induce a $+1$ target
pattern in a confluent layer of NPCs grown on them (phase contrast image at bottom right).
(e): Fluorescence image of nuclei of accumulated NPCs in the target core region. Scale bar is 200$\mu m$.
All experimental details can be found in \cite{Zihuiexper}.}
\end{figure}

Our understanding of cell accumulation around the cores of $+1$ defects is also unsatisfactory. 
In recent experiments on neural progenitor cell (NPC) layers grown on patterned substrates with $+1$ topological defects, 
large-scale flows toward defect cores leading to cell accumulation were observed irrespective of the specific type of
defect considered, be they asters, spirals, or targets 
(see \cite{Zihuiexper}, Fig.~\ref{fig1}(d,e), and Fig.~\ref{fig1}(a-c) for schematic $+1$ defect types).
Current theory, however, predicts cell depletion for targets and strongly chiral spirals.
  
Here, we adopt a bottom-up approach to build a theory motivated by these experimental observations on NPC active nematics. 
Following Patelli {\it et al.}~\cite{Patelli2019}, we first study a particle-level model incorporating the basic ingredients at play, 
including an external field standing for the guiding patterns, 
and derive from it a continuous theory. We show that both model and theory account qualitatively for the main experimental
result, {\it i.e.} accumulation of cells at the core of any type of $+1$ defect. 
We then argue that cell accumulation is essentially due to two generally ignored anti-symmetric active forces
that overcome the standard linear force $-\zeta \nabla \cdot {\bm Q}$.
We finally discuss the relevance and consequences of our findings in the context of other cellular active nematics
experiments and previously proposed theories.

The simplest theory used to describe cellular active nematics \cite{Kawaguchi2017,Copenhagen2021}
usually includes  a `force balance' equation 
\begin{equation}
\gamma {\bm v} = {\bm f}^a \equiv -\zeta \nabla \cdot {\bm Q}
\label{simple} 
\end{equation}
with the coefficient $\gamma$ sometimes replaced by an anisotropic friction
$\gamma \to \gamma_0(I-\varepsilon \boldsymbol{Q})$, where the coefficient $0\leq\varepsilon\leq1$ quantifies larger friction perpendicular to cell alignment \cite{Copenhagen2021}.

In the NPC experiments of interest here (see, e.g., Fig.~\ref{fig1}(d,e) and \cite{Zihuiexper}), 
even though shallow, the ridges on the substrate largely impose that
the orientation of the nematic field ${\bm Q}$ is simply given by that of the $+1$ imprinted defect pattern.
There exists a continuous family of $+1$ topological defects parameterized by the tilt angle $\theta_0=\theta-\phi$ 
that their nematic orientation $\theta$ makes with the reference angle $\phi$ of the polar coordinate frame $(r,\phi)$ 
centered at the defect core (Fig.~\ref{fig1}{b}). In nematic systems, $\theta_0\in [-\tfrac{\pi}{2},\tfrac{\pi}{2}]$, asters corresponds to $\theta=\phi$ hence $\theta_0=0$, while targets are defined by $\theta_0=\pm\tfrac{\pi}{2}$.
All other $\theta_0$ values correspond to spiral patterns (Fig.~\ref{fig1}(a-c)).
Writing axisymmetric ${\bm Q}$ fields as ${\bm Q}=S(r) (\begin{smallmatrix} \cos(2\theta) & \sin(2\theta)  \\ \sin(2\theta)  & -\cos(2\theta) \end{smallmatrix})$, the `active force' ${\bm f}^a$ in \eqref{simple} reads~\cite{SUPP}
\begin{equation}
\bm{f}^a = -\zeta [ S'(r) + \tfrac{2S(r)}{r}] (\cos 2\theta_0 \,\hat{\bm{e}}_r + \sin 2\theta_0 \,\hat{\bm{e}}_\phi),
\end{equation}
where $S'(r)= \frac{dS(r)}{dr}$, and $\hat{\bm{e}}_r$ ($\hat{\bm{e}}_\phi$) is the unit vector in the radial (azimuthal) direction. 
If $\zeta>0$ (``extensile'' case), the radial component is inward for $0 \le |\theta_0| < \frac{\pi}{4}$ and outward
for $\frac{\pi}{4} < |\theta_0| \le \frac{\pi}{2}$ since $S'(r) \ge 0$ holds around the core. 
Arrows in Fig.~\ref{fig1}(a-c) represent the direction of force in typical cases.
Accumulation will thus occur at the cores of  $+1$ defects with $|\theta_0| \le \frac{\pi}{4}$, but
the cores of spirals with $|\theta_0| \ge \frac{\pi}{4}$ and target cores will be depleted, in contradiction 
with the experiments, which found cell accumulation in all cases. 
We further note that anisotropic friction is insufficient to make the target attractive~\cite{SUPP}. 

We now turn to describing the collective dynamics reported in layers of NPCs by a simple particle-level model.
When confluent, these cells take elongated shapes, giving rise to local nematic order, and move along their body axis, stochastically reversing their velocity at a relatively short period of 1-2h (compared to experiments that last several days) \cite{Kawaguchi2017}. These are the ingredients of the Vicsek-style model
for dense active nematics introduced and studied in \cite{Patelli2019}, that we complement here by some external field inducing $+1$ topological defects.
The positions $\bm{r}_i$ and polarity angle $\theta_i$ of point particles obey the overdamped discrete-time dynamics
\begin{eqnarray}
\bm{r}_{i}^{t+1} &=& \bm{r}_i^t + v_0\, \bm{e}_i^{t+1} ,  \label{eqr} \\
\theta_i^{t+1} &=& \arg \lbrace  \epsilon_i^t \lbrack \langle \text{sgn} ( \bm{e}_i^t \cdot \bm{e}_j^t ) \bm{e}_j^t  \rangle_j 
+ \eta \chi_i^t \nonumber \\
&&+ C_p \,\text{sgn} ( \bm{e}_i^t \cdot \bm{e}_i^p ) \bm{e}_i^p \rbrack + C_r \langle \hat{r}_{ji}^t\rangle_j \rbrace   \label{eqtheta}
\end{eqnarray}
where ${\bm e}_i$ is the unit vector with orientation $\theta_i$, $v_0$ is the self-propulsion force,
 $\epsilon_i^t$ is a sign reversing with probability $k$ at each timestep,
 and $\eta$ is the strength of the white uniform angular noise $\chi_i^t \in [-\frac{\pi}{2}, \frac{\pi}{2}]$.
 Averages $\langle \cdot \rangle_j$ in interaction terms are taken over neighboring particles $j$ within unit distance. 
 The first stands for nematic alignment of polarities, and includes the central particle $i$. The second interaction term, of 
 strength $C_r$, codes for soft repulsive torques ($\hat{\bm r}_{ji}^t$ is the unit vector along $\bm{r}_i^t - \bm{r}_j^t$).
 The external field term, of strength $C_p$, represents nematic alignment with the fixed defect pattern: 
 the unit vector $\bm{e}_i^p$ is along the local polar angle $\theta_i^p=\phi_i^t+\theta_0$ where $\phi_i^t$ 
 is the azimuthal angle of the $+1$ defect pattern at location ${\bm r}_i^t$. 
 
\begin{figure}[t!]
\includegraphics[keepaspectratio,width=\columnwidth]{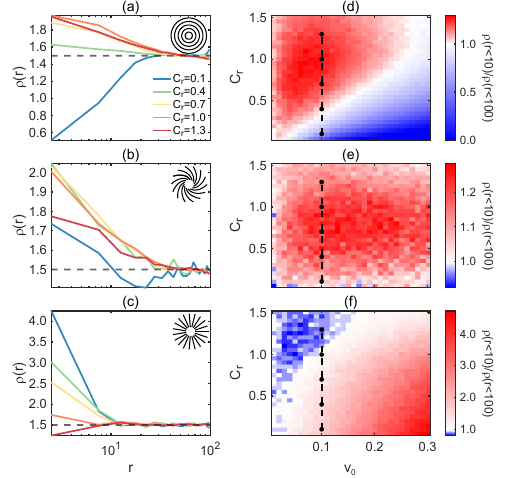}
\caption{\label{fig2} Particle-level model (\ref{eqr},\ref{eqtheta}) in a 
square of linear size $L=200$ with $+1$ defect patterns applied in a central disk of radius $10$
(steady-state results, parameters $\rho_0=1.5, k=0.5, \eta=0.03, C_p = 0.1$).
Left: radial density profiles $\rho(r)$ obtained for $v_0 = 0.1$ at various repulsion strengths $C_r$.
Right: phase diagrams in the $(v_0,C_r)$ plane showing in color the relative accumulation or depletion in the central area
(points along the dashed lines indicate values used in (a-c)).
(a,d): target. (b,e): spiral with $\theta_0=\tfrac{\pi}{4}$. (c,f) aster.}
\end{figure}

This model (with $C_p=0$) was studied at fixed, moderate, repulsion strength $C_r$ in \cite{Patelli2019}.
In order to single out the effects of the imposed $+1$ defect field, we ran simulations 
in the region where the homogeneous nematic (HN) ordered state is stable~\cite{Patelli2019}. 
Applying various $+1$ defect patterns, varying the field strength, we found that, typically, values as low as $C_p = 0.1$ insure that nematic orientation follows the imposed pattern. 
Figure~\ref{fig2}(a-c) shows steady-state radial number density profiles $\rho(r)$ obtained for $C_p = 0.1$ 
For all types of $+1$ defects, particle accumulation is observed in the central region, provided $C_r$ is neither too small nor too large.
We scanned various parameter planes and always found large regions where accumulation occurs. 
Results for the $(v_0, C_r)$ parameter plane are shown in Fig.~\ref{fig2}(d-f). 
(Complementary results can be found in Appendix~A, Fig.~\ref{figEM1}).

Our particle-level model is thus able to account, at least qualitatively, for the systematic accumulation of cells observed
in the NPC experiments.  We now follow \cite{Patelli2019}, and derive a continuous theory from it.
(Details about this derivation can be found in Appendix~B and \cite{SUPP}.) 
The starting point is a  Boltzmann equation governing the one-body distribution function $f(\bm{r}, \theta, t)$ 
that incorporates a field term inducing nematic alignment with the local orientation $\theta_p$ of the imposed pattern:
\begin{eqnarray}
&\partial_t f + v_0 \bm{e}(\theta) \cdot \nabla f + C_p\partial_{\theta} [\sin 2(\theta_p - \theta) f ] = I_{col}[f] \nonumber \\
&+ \bigl [\langle f(\theta - \psi) \rangle_{\psi} - f(\theta) \bigr] + a \bigl [f(\theta + \pi) - f(\theta) \bigr] , \label{eqB}
\end{eqnarray}
where  $\bm{e}(\theta)$ is a unit vector along $\theta$, $I_{col}$ is the collision integral coding for alignment and repulsion, 
 $a$ is the velocity reversal rate, $\langle \cdot \rangle_{\psi}$ is the average over a noise distribution $P(\psi)$ 
 (with rms $\eta$). 
This equation is then truncated and closed using a scaling ansatz.
The result is a set of coupled equations governing the first three angular Fourier modes 
$f_k(\bm{r}, t) = \int_{-\pi}^{\pi} f(\bm{r}, \theta, t)  e^{ik\theta} d\theta$ of $f$:
\begin{eqnarray}
\partial_{t} f_0 &=& -\tfrac{1}{2} v_0 (\triangledown^* f_1 +\triangledown f_1^*), \label{eqrho} \\
\partial_{t} f_1 &=& (-\alpha - \beta |f_2|^2)f_1 + \sigma f_1^* f_2 - \pi_0 \triangledown \rho - \zeta \triangledown^* f_2  \nonumber \\
&+& \lambda (  f_1 \triangledown^* f_1 +  f_1 \triangledown f_1^* -f_2 \triangledown^* \rho ) +  \lambda_3 f_1^* \triangledown f_1 \nonumber \\
&+& \gamma_2 f_2 \triangledown f_2^* + \gamma_1 f_2^* \triangledown f_2 + \tfrac{1}{2}C_p e^{2i \theta_p}f^*_1, \label{eqf1} \\
\partial_{t} f_2 &=& (\mu + \tau |f_1|^2 -\xi |f_2|^2) f_2  + \nu \triangledown\triangledown^* f_2 - \pi_1 \triangledown f_1 \nonumber \\
&+&  \chi_1 \triangledown^* (f_1 f_2) + \chi_2 f_2 \triangledown^* f_1 + \chi_3 f_2 \triangledown f^*_1  + \omega f_1^2 \nonumber \\ 
&+& \kappa_1 f^*_1 \triangledown f_2 + \kappa_2 f_1 \triangledown \rho + C_p e^{2i \theta_p}\rho \,, \label{eqf2}
\end{eqnarray}
where $\triangledown \equiv \partial_x + i\partial_y$ and  $\triangledown^* \equiv \partial_x - i\partial_y$. 
While $f_0$ is real and identical to the density field $\rho$, $f_1$ and $f_2$ are complex-valued, representing 
density-weighted polarity/velocity and nematic order fields 
\begin{equation}
 \bm{w} \equiv \frac{1}{v_0} \rho  \bm{v} = \left(
\begin{array}{r}
\mathcal{R}f_1 \\
\mathcal{I}f_1   
\end{array}
\right), \quad
 \tilde{\bm{Q}} \equiv \rho \bm{Q} =
\left(
\begin{array}{rr}
\mathcal{R}f_2 &  \mathcal{I}f_2 \\
\mathcal{I}f_2  & - \mathcal{R}f_2  
\end{array}
\right). \nonumber
\end{equation}
All coefficients in Eqs.~(\ref{eqrho}-\ref{eqf2}) depend on the microscopic-level parameters 
(see \cite{SUPP} for explicit expressions).
These equations are identical to those derived in \cite{Patelli2019} except for the last term of \eqref{eqf1} and \eqref{eqf2}
which involves the external patterned field.

Equations~(\ref{eqrho}-\ref{eqf2}) can be simulated as is, and their solutions have been shown 
to be qualitatively faithful to the particle-level model (in the absence of external field) \cite{Patelli2019}. 
We have run simulations with imposed $+1$ defect patterns, 
again using parameters mostly in the stable HN ordered state region, 
and found that accumulation at defect cores occurs even for small $C_p$ values. 
Yet, with their many terms, Eqs.~(\ref{eqrho}-\ref{eqf2}) do not easily allow a deeper understanding of 
the features at the origin of accumulation at $+1$ defect cores. 
They can be simplified rather straightforwardly. First of all, as seen in simulations, 
$|f_1|$ is always much smaller than $|f_2|$ in the regimes under
consideration here where nematic order is well developed almost everywhere. 
This suggests to neglect all terms containing $f_1$ twice.
We also found, in simulations with an imposed field pattern, 
that canceling the $f_2\triangledown^*\rho$ term in \eqref{eqf1} and the $\kappa_1$ and $\kappa_2$ terms in \eqref{eqf2} does not impact much the behavior of the system.
Our simplified Eqs.~(\ref{eqrho}-\ref{eqf2}) thus read, using tenso-vectorial notations:
\begin{eqnarray}
\partial_t \rho &=& - v_0 \nabla \cdot \bm{w} , \label{eqrho2}\\
\partial_t \bm{w} &=& [-\alpha - \beta \tilde{S}^2 +\sigma \bm{\tilde{Q}} + \tfrac{1}{2}C_p \bm{Q}^p] \bm{w} - \zeta \nabla \cdot \bm{\tilde{Q}}   \\
&+& \gamma_2 \bm{\tilde{Q}} (\nabla \cdot \bm{\tilde{Q}}) + \gamma_1 [(\bm{\tilde{Q}} \cdot  \nabla) \bm{\tilde{Q}}]^T - 
\pi_0 \nabla \rho \;, \nonumber \label{eqw} \\
\partial_t \bm{\tilde{Q}} &=& [\mu - \xi  \tilde{S}^2]\bm{\tilde{Q}} + \nu \nabla^2 \bm{\tilde{Q}} + (\chi_1 \!+\! \chi_2 \!+\!\chi_3)\bm{\tilde{Q}} (\nabla \cdot \bm{w}) \nonumber \\
&+& (\chi_3 -\chi_2-\chi_1)( \bm{\Omega} \cdot \bm{\tilde{Q}} - \bm{\tilde{Q}} \cdot  \bm{\Omega}) - 2\pi_1 \bm{E} \nonumber \\
&+&  \chi_1 [(\bm{w}\cdot \nabla) \bm{\tilde{Q}} + \bm{\mho} \cdot \bm{\tilde{Q}}] + C_p \rho \,\bm{Q}^p  , \label{eqQ}
\end{eqnarray}
where $\tilde{\bm Q}^p$ is the nematic field of the imposed pattern, $\tilde{S}=\rho S$,
$\bm{E}$ is the symmetric strain rate-like tensor defined by $E_{ij} = \frac{1}{2}(\partial_i w_j + \partial_j w_i - \delta_{ij}\nabla \cdot \bm{w})$, 
$\bm{\Omega}$ is the antisymmetric rate of rotation tensor $\Omega_{ij} = \frac{1}{2}(\partial_i w_j - \partial_j w_i)$, 
and $\mho_{ij} = w_j\partial_i - w_i \partial_j$.
Note that Eq.~\eqref{eqQ} is similar to the Beris–Edwards equation~\cite{deGennes1993, beris1994thermodynamics,marenduzzo2007steady,doostmohammadi2018}  

\begin{figure}[b!]
\includegraphics[keepaspectratio, width=\columnwidth]{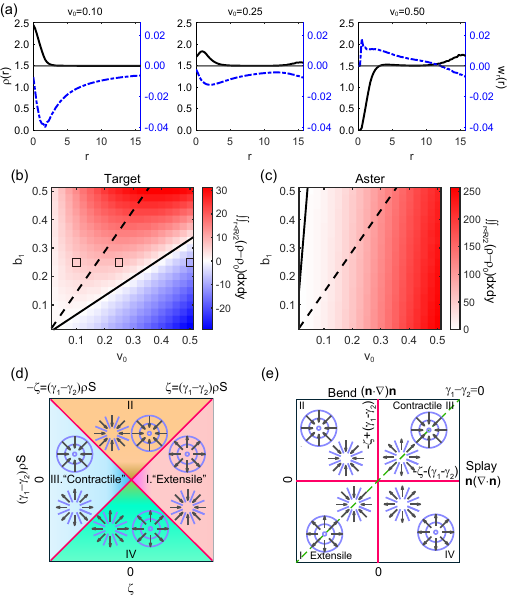}
\caption{\label{fig3}
(a-c) Simulations of Eqs.~(\ref{eqrho2}-\ref{eqQ}) in a square domain of linear size $L=32$ with periodic boundary conditions. 
A patterning field of strength $C_p = 0.1$ is applied in the central disk of radius 16 ($a=0.4$, $\eta=0.2$, $\nu=0.2$, $\rho_0=1.5$, pseudo-spectral method, explicit Euler time-stepping, $dx=0.25$, $dt=0.01$, only steady-state results are shown).
(a): Instantaneous, azimuthally-averaged, radial density $\rho(r)$ (left y-axis) and density-weighted radial velocity $w_r(r)$ (right y-axis) profiles observed when a target pattern is applied with $v_0 = 0.1, 0.25$, and $0.5$ from left to right ($b_1=0.25$, square symbols in (b)). 
 (The `irregularities' seen for $r\gtrsim12$ are due to the self-interacting external boundary of the patterned region.)
 (b,c): Phase diagrams in the $(v_0,b_1)$ plane showing accumulation or depletion
((b): target pattern as in (a); (c): aster pattern).
The color scale codes for the mass surplus (red) or deficit (blue) with respect to the mean density $\rho_0$ measured in a central disk of radius 8 outside which the local density is very near $\rho_0$.
The solid lines correspond to the analytical prediction 
$\pm\zeta - (\gamma_1 - \gamma_2) \sqrt{\mu/\xi}=0$ (see main text), while the dashed lines indicate $\zeta=0$.
(d): Classification of active nematics in the $[\zeta, (\gamma_1-\gamma_2)\rho S]$ plane.  
In each region delimited, the direction of radial flow is indicated for targets and asters.
(e): Classification of active nematics in the (splay, bend) plane when $S=1$ and $\rho$ is uniform. Splay mode amplitude is proportional to $-\zeta-(\gamma_1-\gamma_2)$, bend mode amplitude is proportional to $-\zeta + (\gamma_1-\gamma_2)$.}
\end{figure}

Numerical simulations of Eqs.~(\ref{eqrho2}-\ref{eqQ}) and of Eqs.~(\ref{eqrho}-\ref{eqf2}), 
performed at similar parameters, yield very similar results~\footnote{One quantitative difference
is that a smaller $C_p$ value is sufficient, for the simplified equations, to fully impose the external pattern 
and in particular to avoid that the $+1$ defect splits into two $+\tfrac{1}{2}$ ones.}.
We now show features of the typical axisymmetric steady-state solutions of Eqs.~(\ref{eqrho2}-\ref{eqQ}) 
when imposing weak patterns ($C_p = 0.1$). 
In Fig.~\ref{fig3}(a), where a target pattern is induced, radial density and velocity profiles reveal that inward flows and accumulation in the core region are only replaced by outward flow  and depletion for large enough nominal speed $v_0$.
Figure~\ref{fig3}(b,c) display phase diagrams in a parameter plane equivalent to that of Fig.~\ref{fig1}(d-f) for both 
target and aster patterns. 
Showing large regions of accumulation near defect cores, they are very similar to those obtained at particle level.  

To obtain some analytical insights and identify the roles of active forces around $+1$ defect, we 
make further simplifications. 
First of all, we neglect
the $\rho$ dependence of coefficients ($\alpha, \mu, \pi_0, \pi_1,$ and $\zeta$ are impacted). 
 In the steady state, Eq.~\eqref{eqw} yields
\begin{eqnarray}
&& \bm{w} = {\Gamma(\tilde{\bm{Q}},{\bm Q}^p)}^{-1}(\bm{f}^a - \pi_0 \nabla \rho) \label{eqwsol}\\ 
&\!\!\!\!\!\!\!\! {\rm with}\;\;& \bm{f}^a = (- \zeta + \gamma_2 \bm{\tilde{Q}}) \nabla \cdot \bm{\tilde{Q}} + \gamma_1 [(\bm{\tilde{Q}} \cdot \nabla) \bm{\tilde{Q}}]^T , \\
&&\Gamma(\tilde{\bm{Q}},{\bm Q}^p)= \alpha +\beta \tilde{S}^2 -\sigma \tilde{\bm{Q}}- \tfrac{1}{2}C_p \bm{Q}^p.
\end{eqnarray}
When $\tilde{\bm{Q}}$ is well aligned with the external field in the steady state and
$\bm{w}$ is negligible compared with $\tilde{\bm{Q}}$ in Eq.~\eqref{eqQ}, $[\mu - \xi  \tilde{S}^2]\bm{\tilde{Q}}+C_p \bm{Q}^p \rho=0$ holds. Writing 
$\tilde{\bm Q} = \rho S {\bm Q}^p$ yields $[\mu - \xi  \rho^2 S^2]S +C_p =0$.
Solving for $\rho S$ assuming small deviations from $\sqrt{\mu/\xi}$ ($\mu, \xi>0$), we obtain 
$\rho S \sim \sqrt{\mu/\xi} + \frac{\rho C_p}{2\mu} +\mathcal{O}(C_p^2)$. 
Injecting this expression in $\Gamma$ yields  $\Gamma(\tilde{\bm{Q}},{\bm Q}^p)= \Gamma(\tilde{\bm{Q}}) =\gamma_0 (\bm{I} - \varepsilon \tilde{\bm{Q}}).$
where $\gamma_0 = \alpha + \beta \tilde{S}^2>0$ and $\varepsilon =[\sigma + C_p/(2\rho S)] /\gamma_0$. ($\alpha>0, \beta>0$)
In \cite{SUPP}, we argue that, for axisymmetric solutions, the anisotropy coefficient $\varepsilon$ 
(in $\Gamma(\tilde{\bm{Q}})$) 
cannot reverse the sign of the radial component of ${\bm w}$. Thus, this radial flow vanishes (steady-state condition) 
when $f_r= \pi_0 d\rho/dr$ where $f_r$ is the radial component of ${\bm f}^a$.
Consequently, accumulation or depletion, governed by the sign of $d\rho/dr$, is decided by the sign of $f_r$.

We also show in \cite{SUPP} that $f_r$, when a target or aster pattern is imposed, can be approximated by:
\begin{equation}
    f_r(r) \sim  \bigl[ \pm \zeta - (\gamma_1 - \gamma_2)\rho S \bigr] \frac{2\rho S}{r},  \label{eqfr}
\end{equation}
where the $+$ sign is for a target, and the $-$ sign is for an aster. 
We thus conclude that under our approximations, the accumulation-depletion transition is given by $f_r=0$. 
On the transition line, we can safely assume that $\rho = \rho_0$.
For weak fields ($C_p\ll1$) and away from the defect core, this gives 
$\pm\zeta - (\gamma_1 - \gamma_2) \sqrt{\mu(\rho_0)/\xi}=0$. 
The corresponding lines, indicated by the solid lines in the phase diagrams of Fig.~\ref{fig3}(b,c), match very well the 
numerical results for Eqs.(9-11) presented there. On the other hand, the transition lines $\zeta=0$ (the dashed lines) given by considering only the linear force 
($\gamma_1=\gamma_2=0$), are unable to account for the accumulation-depletion transition.

Our results suggest a novel classification of cellular active nematics. 
From Eq.~\eqref{eqfr}, we see that $\gamma_1-\gamma_2$  is a crucial quantity to decide whether $+1$ defects
will lead to accumulation or depletion. This comes in addition to the sign and value of $\zeta$, the coefficient of the standard
linear force term. The plane $(\zeta,\gamma_1-\gamma_2)$  can be divided into 4 quadrants classifying various responses to
the presence of $+1$ defects. Sketched in Fig.~\ref{fig3}(d), it shows how the traditional extensile-contractile dichotomy, governed by the sign of $\zeta$, is unable to account alone for the phenomena at stake.
Our bottom-up approach yields expressions for these coefficients, and shows that
$\gamma_2 < 0$ holds for the systems with repulsive interactions, while $\gamma_1 > 0$ in most cases (see Appendix~B), 
so that $\gamma_1-\gamma_2>0$ for the NPC layers of interest here, which appear to be located in zone~II.
We confirmed further the validity of our conclusion by varying $\zeta$ and $\gamma_1 - \gamma_2$ independently of the coefficients derived from the particle model, considering also the case of $-1$ defects (see Appendix~C).

Finally, to grasp the intuitive meaning of the $\gamma_1$ and $\gamma_2$ terms, we rewrote them 
assuming $S=1$ and uniform density. They can then be recombined, together with the usual $\zeta$
active stress, into classic bend and splay terms:
\begin{eqnarray}
 &-\zeta \nabla \cdot \tilde{\bm{Q}} + \gamma_1 [(\tilde{\bm{Q}} \cdot \nabla)\tilde{\bm{Q}}]^T + \gamma_2 \tilde{\bm{Q}} (\nabla \cdot \tilde{\bm{Q}}) \nonumber \\
    \simeq& (A+B) (\bm{n}\cdot \nabla)\bm{n} + (A-B) \bm{n} (\nabla \cdot \bm{n}) ,
\end{eqnarray}
where $A = -2\zeta$ and $B = 2(\gamma_1 - \gamma_2)$, and $\bm{n}=(n_x, n_y)$ is the director field (see \cite{SUPP}). 
The classification diagram (Fig.~\ref{fig3}(d)) can then be re-expressed as in Fig.~\ref{fig3}(e). 
In this diagram, the horizontal axis is pure splay, and the vertical axis is pure bend.
 As long as the anti-symmetric term exists, $\gamma_1-\gamma_2\neq 0$, the one-constant approximation is not valid. 
Thus, there are always situations where the behavior near integer defects cannot be classified extensile (I) or contractile (III).

To summarize, we have demonstrated the importance of the two generally ignored anti-symmetric forces
and shown that the conventional active stress term alone cannot account for the 
accumulation towards the core of $+1$ topological defects in cellular active nematics such as the NPC layers of interest here.
All in all, we believe our results point to the importance of nonlinear terms often absent from theories of 2D active nematics systems.

\acknowledgments
We are grateful to Aurelio Patelli, Beno\^{\i}t Mahault, Carles Blanch-Mercader, Eric Bertin, Jacques Prost, and Sriram Ramaswamy for insightful discussions.
We acknowledge support from NSFC (12225410, 12074243) for H.Z. and (12174254, 12250710131) for M.S.

\bibliographystyle{apsrev4-2}
\bibliography{main}

\begin{thebibliography}{45}%
\makeatletter
\providecommand \@ifxundefined [1]{%
 \@ifx{#1\undefined}
}%
\providecommand \@ifnum [1]{%
 \ifnum #1\expandafter \@firstoftwo
 \else \expandafter \@secondoftwo
 \fi
}%
\providecommand \@ifx [1]{%
 \ifx #1\expandafter \@firstoftwo
 \else \expandafter \@secondoftwo
 \fi
}%
\providecommand \natexlab [1]{#1}%
\providecommand \enquote  [1]{``#1''}%
\providecommand \bibnamefont  [1]{#1}%
\providecommand \bibfnamefont [1]{#1}%
\providecommand \citenamefont [1]{#1}%
\providecommand \href@noop [0]{\@secondoftwo}%
\providecommand \href [0]{\begingroup \@sanitize@url \@href}%
\providecommand \@href[1]{\@@startlink{#1}\@@href}%
\providecommand \@@href[1]{\endgroup#1\@@endlink}%
\providecommand \@sanitize@url [0]{\catcode `\\12\catcode `\$12\catcode `\&12\catcode `\#12\catcode `\^12\catcode `\_12\catcode `\%12\relax}%
\providecommand \@@startlink[1]{}%
\providecommand \@@endlink[0]{}%
\providecommand \url  [0]{\begingroup\@sanitize@url \@url }%
\providecommand \@url [1]{\endgroup\@href {#1}{\urlprefix }}%
\providecommand \urlprefix  [0]{URL }%
\providecommand \Eprint [0]{\href }%
\providecommand \doibase [0]{https://doi.org/}%
\providecommand \selectlanguage [0]{\@gobble}%
\providecommand \bibinfo  [0]{\@secondoftwo}%
\providecommand \bibfield  [0]{\@secondoftwo}%
\providecommand \translation [1]{[#1]}%
\providecommand \BibitemOpen [0]{}%
\providecommand \bibitemStop [0]{}%
\providecommand \bibitemNoStop [0]{.\EOS\space}%
\providecommand \EOS [0]{\spacefactor3000\relax}%
\providecommand \BibitemShut  [1]{\csname bibitem#1\endcsname}%
\let\auto@bib@innerbib\@empty
\bibitem [{\citenamefont {Kruse}\ \emph {et~al.}(2004)\citenamefont {Kruse}, \citenamefont {Joanny}, \citenamefont {J{\"{u}}licher}, \citenamefont {Prost},\ and\ \citenamefont {Sekimoto}}]{Prost2004}%
  \BibitemOpen
  \bibfield  {author} {\bibinfo {author} {\bibfnamefont {K.}~\bibnamefont {Kruse}}, \bibinfo {author} {\bibfnamefont {J.~F.}\ \bibnamefont {Joanny}}, \bibinfo {author} {\bibfnamefont {F.}~\bibnamefont {J{\"{u}}licher}}, \bibinfo {author} {\bibfnamefont {J.}~\bibnamefont {Prost}},\ and\ \bibinfo {author} {\bibfnamefont {K.}~\bibnamefont {Sekimoto}},\ }\href {https://doi.org/10.1103/PhysRevLett.92.078101} {\bibfield  {journal} {\bibinfo  {journal} {Physical Review Letters}\ }\textbf {\bibinfo {volume} {92}},\ \bibinfo {pages} {1} (\bibinfo {year} {2004})}\BibitemShut {NoStop}%
\bibitem [{\citenamefont {Marchetti}\ \emph {et~al.}(2013)\citenamefont {Marchetti}, \citenamefont {Joanny}, \citenamefont {Ramaswamy}, \citenamefont {Liverpool}, \citenamefont {Prost}, \citenamefont {Rao},\ and\ \citenamefont {Simha}}]{Marchetti2013}%
  \BibitemOpen
  \bibfield  {author} {\bibinfo {author} {\bibfnamefont {M.~C.}\ \bibnamefont {Marchetti}}, \bibinfo {author} {\bibfnamefont {J.~F.}\ \bibnamefont {Joanny}}, \bibinfo {author} {\bibfnamefont {S.}~\bibnamefont {Ramaswamy}}, \bibinfo {author} {\bibfnamefont {T.~B.}\ \bibnamefont {Liverpool}}, \bibinfo {author} {\bibfnamefont {J.}~\bibnamefont {Prost}}, \bibinfo {author} {\bibfnamefont {M.}~\bibnamefont {Rao}},\ and\ \bibinfo {author} {\bibfnamefont {R.~A.}\ \bibnamefont {Simha}},\ }\href {https://doi.org/10.1103/RevModPhys.85.1143} {\bibfield  {journal} {\bibinfo  {journal} {Reviews of Modern Physics}\ }\textbf {\bibinfo {volume} {85}},\ \bibinfo {pages} {1143} (\bibinfo {year} {2013})}\BibitemShut {NoStop}%
\bibitem [{\citenamefont {Shankar}\ \emph {et~al.}(2022)\citenamefont {Shankar}, \citenamefont {Souslov}, \citenamefont {Bowick}, \citenamefont {Marchetti},\ and\ \citenamefont {Vitelli}}]{Shankar2022}%
  \BibitemOpen
  \bibfield  {author} {\bibinfo {author} {\bibfnamefont {S.}~\bibnamefont {Shankar}}, \bibinfo {author} {\bibfnamefont {A.}~\bibnamefont {Souslov}}, \bibinfo {author} {\bibfnamefont {M.~J.}\ \bibnamefont {Bowick}}, \bibinfo {author} {\bibfnamefont {M.~C.}\ \bibnamefont {Marchetti}},\ and\ \bibinfo {author} {\bibfnamefont {V.}~\bibnamefont {Vitelli}},\ }\href {https://doi.org/10.1038/s42254-022-00445-3} {\bibfield  {journal} {\bibinfo  {journal} {Nature Reviews Physics}\ }\textbf {\bibinfo {volume} {4}},\ \bibinfo {pages} {380} (\bibinfo {year} {2022})}\BibitemShut {NoStop}%
\bibitem [{\citenamefont {Gompper}\ \emph {et~al.}(2020)\citenamefont {Gompper}, \citenamefont {Winkler}, \citenamefont {Speck}, \citenamefont {Solon}, \citenamefont {Nardini}, \citenamefont {Peruani}, \citenamefont {L{\"o}wen}, \citenamefont {Golestanian}, \citenamefont {Kaupp}, \citenamefont {Alvarez} \emph {et~al.}}]{gompper20202020}%
  \BibitemOpen
  \bibfield  {author} {\bibinfo {author} {\bibfnamefont {G.}~\bibnamefont {Gompper}}, \bibinfo {author} {\bibfnamefont {R.~G.}\ \bibnamefont {Winkler}}, \bibinfo {author} {\bibfnamefont {T.}~\bibnamefont {Speck}}, \bibinfo {author} {\bibfnamefont {A.}~\bibnamefont {Solon}}, \bibinfo {author} {\bibfnamefont {C.}~\bibnamefont {Nardini}}, \bibinfo {author} {\bibfnamefont {F.}~\bibnamefont {Peruani}}, \bibinfo {author} {\bibfnamefont {H.}~\bibnamefont {L{\"o}wen}}, \bibinfo {author} {\bibfnamefont {R.}~\bibnamefont {Golestanian}}, \bibinfo {author} {\bibfnamefont {U.~B.}\ \bibnamefont {Kaupp}}, \bibinfo {author} {\bibfnamefont {L.}~\bibnamefont {Alvarez}}, \emph {et~al.},\ }\href@noop {} {\bibfield  {journal} {\bibinfo  {journal} {Journal of Physics: Condensed Matter}\ }\textbf {\bibinfo {volume} {32}},\ \bibinfo {pages} {193001} (\bibinfo {year} {2020})}\BibitemShut {NoStop}%
\bibitem [{\citenamefont {Duclos}\ \emph {et~al.}(2017)\citenamefont {Duclos}, \citenamefont {Erlenk{\"{a}}mper}, \citenamefont {Joanny},\ and\ \citenamefont {Silberzan}}]{Duclos2017}%
  \BibitemOpen
  \bibfield  {author} {\bibinfo {author} {\bibfnamefont {G.}~\bibnamefont {Duclos}}, \bibinfo {author} {\bibfnamefont {C.}~\bibnamefont {Erlenk{\"{a}}mper}}, \bibinfo {author} {\bibfnamefont {J.~F.}\ \bibnamefont {Joanny}},\ and\ \bibinfo {author} {\bibfnamefont {P.}~\bibnamefont {Silberzan}},\ }\href {https://doi.org/10.1038/nphys3876} {\bibfield  {journal} {\bibinfo  {journal} {Nature Physics}\ }\textbf {\bibinfo {volume} {13}},\ \bibinfo {pages} {58} (\bibinfo {year} {2017})}\BibitemShut {NoStop}%
\bibitem [{\citenamefont {Copenhagen}\ \emph {et~al.}(2021)\citenamefont {Copenhagen}, \citenamefont {Alert}, \citenamefont {Wingreen},\ and\ \citenamefont {Shaevitz}}]{Copenhagen2021}%
  \BibitemOpen
  \bibfield  {author} {\bibinfo {author} {\bibfnamefont {K.}~\bibnamefont {Copenhagen}}, \bibinfo {author} {\bibfnamefont {R.}~\bibnamefont {Alert}}, \bibinfo {author} {\bibfnamefont {N.~S.}\ \bibnamefont {Wingreen}},\ and\ \bibinfo {author} {\bibfnamefont {J.~W.}\ \bibnamefont {Shaevitz}},\ }\href {https://doi.org/10.1038/s41567-020-01056-4} {\bibfield  {journal} {\bibinfo  {journal} {Nature Physics}\ }\textbf {\bibinfo {volume} {17}},\ \bibinfo {pages} {211} (\bibinfo {year} {2021})}\BibitemShut {NoStop}%
\bibitem [{\citenamefont {Maroudas-Sacks}\ \emph {et~al.}(2021)\citenamefont {Maroudas-Sacks}, \citenamefont {Garion}, \citenamefont {Shani-Zerbib}, \citenamefont {Livshits}, \citenamefont {Braun},\ and\ \citenamefont {Keren}}]{Maroudas-Sacks2021}%
  \BibitemOpen
  \bibfield  {author} {\bibinfo {author} {\bibfnamefont {Y.}~\bibnamefont {Maroudas-Sacks}}, \bibinfo {author} {\bibfnamefont {L.}~\bibnamefont {Garion}}, \bibinfo {author} {\bibfnamefont {L.}~\bibnamefont {Shani-Zerbib}}, \bibinfo {author} {\bibfnamefont {A.}~\bibnamefont {Livshits}}, \bibinfo {author} {\bibfnamefont {E.}~\bibnamefont {Braun}},\ and\ \bibinfo {author} {\bibfnamefont {K.}~\bibnamefont {Keren}},\ }\href {https://doi.org/10.1038/s41567-020-01083-1} {\bibfield  {journal} {\bibinfo  {journal} {Nature Physics}\ }\textbf {\bibinfo {volume} {17}},\ \bibinfo {pages} {251} (\bibinfo {year} {2021})}\BibitemShut {NoStop}%
\bibitem [{\citenamefont {Duclos}\ \emph {et~al.}(2014)\citenamefont {Duclos}, \citenamefont {Garcia}, \citenamefont {Yevick},\ and\ \citenamefont {Silberzan}}]{duclos2014}%
  \BibitemOpen
  \bibfield  {author} {\bibinfo {author} {\bibfnamefont {G.}~\bibnamefont {Duclos}}, \bibinfo {author} {\bibfnamefont {S.}~\bibnamefont {Garcia}}, \bibinfo {author} {\bibfnamefont {H.}~\bibnamefont {Yevick}},\ and\ \bibinfo {author} {\bibfnamefont {P.}~\bibnamefont {Silberzan}},\ }\href@noop {} {\bibfield  {journal} {\bibinfo  {journal} {Soft matter}\ }\textbf {\bibinfo {volume} {10}},\ \bibinfo {pages} {2346} (\bibinfo {year} {2014})}\BibitemShut {NoStop}%
\bibitem [{\citenamefont {Duclos}\ \emph {et~al.}(2018)\citenamefont {Duclos}, \citenamefont {Blanch-Mercader}, \citenamefont {Yashunsky}, \citenamefont {Salbreux}, \citenamefont {Joanny}, \citenamefont {Prost},\ and\ \citenamefont {Silberzan}}]{Duclos2018}%
  \BibitemOpen
  \bibfield  {author} {\bibinfo {author} {\bibfnamefont {G.}~\bibnamefont {Duclos}}, \bibinfo {author} {\bibfnamefont {C.}~\bibnamefont {Blanch-Mercader}}, \bibinfo {author} {\bibfnamefont {V.}~\bibnamefont {Yashunsky}}, \bibinfo {author} {\bibfnamefont {G.}~\bibnamefont {Salbreux}}, \bibinfo {author} {\bibfnamefont {J.~F.}\ \bibnamefont {Joanny}}, \bibinfo {author} {\bibfnamefont {J.}~\bibnamefont {Prost}},\ and\ \bibinfo {author} {\bibfnamefont {P.}~\bibnamefont {Silberzan}},\ }\href {https://doi.org/10.1038/s41567-018-0099-7} {\bibfield  {journal} {\bibinfo  {journal} {Nature Physics}\ }\textbf {\bibinfo {volume} {14}},\ \bibinfo {pages} {728} (\bibinfo {year} {2018})}\BibitemShut {NoStop}%
\bibitem [{\citenamefont {Sarkar}\ \emph {et~al.}(2023)\citenamefont {Sarkar}, \citenamefont {Yashunsky}, \citenamefont {Br{\'e}zin}, \citenamefont {Blanch~Mercader}, \citenamefont {Aryaksama}, \citenamefont {Lacroix}, \citenamefont {Risler}, \citenamefont {Joanny},\ and\ \citenamefont {Silberzan}}]{sarkar2023crisscross}%
  \BibitemOpen
  \bibfield  {author} {\bibinfo {author} {\bibfnamefont {T.}~\bibnamefont {Sarkar}}, \bibinfo {author} {\bibfnamefont {V.}~\bibnamefont {Yashunsky}}, \bibinfo {author} {\bibfnamefont {L.}~\bibnamefont {Br{\'e}zin}}, \bibinfo {author} {\bibfnamefont {C.}~\bibnamefont {Blanch~Mercader}}, \bibinfo {author} {\bibfnamefont {T.}~\bibnamefont {Aryaksama}}, \bibinfo {author} {\bibfnamefont {M.}~\bibnamefont {Lacroix}}, \bibinfo {author} {\bibfnamefont {T.}~\bibnamefont {Risler}}, \bibinfo {author} {\bibfnamefont {J.-F.}\ \bibnamefont {Joanny}},\ and\ \bibinfo {author} {\bibfnamefont {P.}~\bibnamefont {Silberzan}},\ }\href@noop {} {\bibfield  {journal} {\bibinfo  {journal} {PNAS nexus}\ }\textbf {\bibinfo {volume} {2}},\ \bibinfo {pages} {pgad034} (\bibinfo {year} {2023})}\BibitemShut {NoStop}%
\bibitem [{\citenamefont {Hakim}\ and\ \citenamefont {Silberzan}(2017)}]{hakim2017}%
  \BibitemOpen
  \bibfield  {author} {\bibinfo {author} {\bibfnamefont {V.}~\bibnamefont {Hakim}}\ and\ \bibinfo {author} {\bibfnamefont {P.}~\bibnamefont {Silberzan}},\ }\href@noop {} {\bibfield  {journal} {\bibinfo  {journal} {Reports on Progress in Physics}\ }\textbf {\bibinfo {volume} {80}},\ \bibinfo {pages} {076601} (\bibinfo {year} {2017})}\BibitemShut {NoStop}%
\bibitem [{\citenamefont {Saw}\ \emph {et~al.}(2017)\citenamefont {Saw}, \citenamefont {Doostmohammadi}, \citenamefont {Nier}, \citenamefont {Kocgozlu}, \citenamefont {Thampi}, \citenamefont {Toyama}, \citenamefont {Marcq}, \citenamefont {Lim}, \citenamefont {Yeomans},\ and\ \citenamefont {Ladoux}}]{Saw2017}%
  \BibitemOpen
  \bibfield  {author} {\bibinfo {author} {\bibfnamefont {T.~B.}\ \bibnamefont {Saw}}, \bibinfo {author} {\bibfnamefont {A.}~\bibnamefont {Doostmohammadi}}, \bibinfo {author} {\bibfnamefont {V.}~\bibnamefont {Nier}}, \bibinfo {author} {\bibfnamefont {L.}~\bibnamefont {Kocgozlu}}, \bibinfo {author} {\bibfnamefont {S.}~\bibnamefont {Thampi}}, \bibinfo {author} {\bibfnamefont {Y.}~\bibnamefont {Toyama}}, \bibinfo {author} {\bibfnamefont {P.}~\bibnamefont {Marcq}}, \bibinfo {author} {\bibfnamefont {C.~T.}\ \bibnamefont {Lim}}, \bibinfo {author} {\bibfnamefont {J.~M.}\ \bibnamefont {Yeomans}},\ and\ \bibinfo {author} {\bibfnamefont {B.}~\bibnamefont {Ladoux}},\ }\href {http://dx.doi.org/10.1038/nature21718} {\bibfield  {journal} {\bibinfo  {journal} {Nature}\ }\textbf {\bibinfo {volume} {544}},\ \bibinfo {pages} {212} (\bibinfo {year} {2017})}\BibitemShut {NoStop}%
\bibitem [{\citenamefont {Kawaguchi}\ \emph {et~al.}(2017)\citenamefont {Kawaguchi}, \citenamefont {Kageyama},\ and\ \citenamefont {Sano}}]{Kawaguchi2017}%
  \BibitemOpen
  \bibfield  {author} {\bibinfo {author} {\bibfnamefont {K.}~\bibnamefont {Kawaguchi}}, \bibinfo {author} {\bibfnamefont {R.}~\bibnamefont {Kageyama}},\ and\ \bibinfo {author} {\bibfnamefont {M.}~\bibnamefont {Sano}},\ }\href {https://doi.org/10.1038/nature22321} {\bibfield  {journal} {\bibinfo  {journal} {Nature}\ }\textbf {\bibinfo {volume} {545}},\ \bibinfo {pages} {327} (\bibinfo {year} {2017})}\BibitemShut {NoStop}%
\bibitem [{\citenamefont {Giomi}\ \emph {et~al.}(2022)\citenamefont {Giomi}, \citenamefont {Carenza}, \citenamefont {Eckert},\ and\ \citenamefont {Luca}}]{Hoffmann2022}%
  \BibitemOpen
  \bibfield  {author} {\bibinfo {author} {\bibfnamefont {L.~A.~H.}\ \bibnamefont {Giomi}}, \bibinfo {author} {\bibfnamefont {L.~N.}\ \bibnamefont {Carenza}}, \bibinfo {author} {\bibfnamefont {J.}~\bibnamefont {Eckert}},\ and\ \bibinfo {author} {\bibnamefont {Luca}},\ }\href {https://doi.org/0.1126/sciadv.abk2712} {\bibfield  {journal} {\bibinfo  {journal} {Science Advances}\ }\textbf {\bibinfo {volume} {8}},\ \bibinfo {pages} {eabk2712} (\bibinfo {year} {2022})}\BibitemShut {NoStop}%
\bibitem [{\citenamefont {Ho}\ \emph {et~al.}(2024)\citenamefont {Ho}, \citenamefont {B{\o}e}, \citenamefont {Dysthe},\ and\ \citenamefont {Angheluta}}]{ho2024role}%
  \BibitemOpen
  \bibfield  {author} {\bibinfo {author} {\bibfnamefont {R.~D.}\ \bibnamefont {Ho}}, \bibinfo {author} {\bibfnamefont {S.~O.}\ \bibnamefont {B{\o}e}}, \bibinfo {author} {\bibfnamefont {D.~K.}\ \bibnamefont {Dysthe}},\ and\ \bibinfo {author} {\bibfnamefont {L.}~\bibnamefont {Angheluta}},\ }\href@noop {} {\bibfield  {journal} {\bibinfo  {journal} {Physical Review Research}\ }\textbf {\bibinfo {volume} {6}},\ \bibinfo {pages} {023315} (\bibinfo {year} {2024})}\BibitemShut {NoStop}%
\bibitem [{\citenamefont {Mietke}\ \emph {et~al.}(2019{\natexlab{a}})\citenamefont {Mietke}, \citenamefont {J{\"u}licher},\ and\ \citenamefont {Sbalzarini}}]{mietke2019self}%
  \BibitemOpen
  \bibfield  {author} {\bibinfo {author} {\bibfnamefont {A.}~\bibnamefont {Mietke}}, \bibinfo {author} {\bibfnamefont {F.}~\bibnamefont {J{\"u}licher}},\ and\ \bibinfo {author} {\bibfnamefont {I.~F.}\ \bibnamefont {Sbalzarini}},\ }\href@noop {} {\bibfield  {journal} {\bibinfo  {journal} {Proceedings of the National Academy of Sciences}\ }\textbf {\bibinfo {volume} {116}},\ \bibinfo {pages} {29} (\bibinfo {year} {2019}{\natexlab{a}})}\BibitemShut {NoStop}%
\bibitem [{\citenamefont {Mietke}\ \emph {et~al.}(2019{\natexlab{b}})\citenamefont {Mietke}, \citenamefont {Jemseena}, \citenamefont {Kumar}, \citenamefont {Sbalzarini},\ and\ \citenamefont {J{\"u}licher}}]{mietke2019minimal}%
  \BibitemOpen
  \bibfield  {author} {\bibinfo {author} {\bibfnamefont {A.}~\bibnamefont {Mietke}}, \bibinfo {author} {\bibfnamefont {V.}~\bibnamefont {Jemseena}}, \bibinfo {author} {\bibfnamefont {K.~V.}\ \bibnamefont {Kumar}}, \bibinfo {author} {\bibfnamefont {I.~F.}\ \bibnamefont {Sbalzarini}},\ and\ \bibinfo {author} {\bibfnamefont {F.}~\bibnamefont {J{\"u}licher}},\ }\href@noop {} {\bibfield  {journal} {\bibinfo  {journal} {Physical review letters}\ }\textbf {\bibinfo {volume} {123}},\ \bibinfo {pages} {188101} (\bibinfo {year} {2019}{\natexlab{b}})}\BibitemShut {NoStop}%
\bibitem [{\citenamefont {Hamant}\ \emph {et~al.}(2008)\citenamefont {Hamant}, \citenamefont {Heisler}, \citenamefont {Jonsson}, \citenamefont {Krupinski}, \citenamefont {Uyttewaal}, \citenamefont {Bokov}, \citenamefont {Corson}, \citenamefont {Sahlin}, \citenamefont {Boudaoud}, \citenamefont {Meyerowitz} \emph {et~al.}}]{hamant2008}%
  \BibitemOpen
  \bibfield  {author} {\bibinfo {author} {\bibfnamefont {O.}~\bibnamefont {Hamant}}, \bibinfo {author} {\bibfnamefont {M.~G.}\ \bibnamefont {Heisler}}, \bibinfo {author} {\bibfnamefont {H.}~\bibnamefont {Jonsson}}, \bibinfo {author} {\bibfnamefont {P.}~\bibnamefont {Krupinski}}, \bibinfo {author} {\bibfnamefont {M.}~\bibnamefont {Uyttewaal}}, \bibinfo {author} {\bibfnamefont {P.}~\bibnamefont {Bokov}}, \bibinfo {author} {\bibfnamefont {F.}~\bibnamefont {Corson}}, \bibinfo {author} {\bibfnamefont {P.}~\bibnamefont {Sahlin}}, \bibinfo {author} {\bibfnamefont {A.}~\bibnamefont {Boudaoud}}, \bibinfo {author} {\bibfnamefont {E.~M.}\ \bibnamefont {Meyerowitz}}, \emph {et~al.},\ }\href@noop {} {\bibfield  {journal} {\bibinfo  {journal} {Science}\ }\textbf {\bibinfo {volume} {322}},\ \bibinfo {pages} {1650} (\bibinfo {year} {2008})}\BibitemShut {NoStop}%
\bibitem [{\citenamefont {Chandrasekhar}(1992)}]{Schandrasekhar2013}%
  \BibitemOpen
  \bibfield  {author} {\bibinfo {author} {\bibfnamefont {S.}~\bibnamefont {Chandrasekhar}},\ }\href@noop {} {\emph {\bibinfo {title} {Liquid Crystals}}}\ (\bibinfo  {publisher} {Cambridge University Press},\ \bibinfo {year} {1992})\BibitemShut {NoStop}%
\bibitem [{\citenamefont {Guillamat}\ \emph {et~al.}(2022)\citenamefont {Guillamat}, \citenamefont {Blanch-Mercader}, \citenamefont {Pernollet}, \citenamefont {Kruse},\ and\ \citenamefont {Roux}}]{Guillamat}%
  \BibitemOpen
  \bibfield  {author} {\bibinfo {author} {\bibfnamefont {P.}~\bibnamefont {Guillamat}}, \bibinfo {author} {\bibfnamefont {C.}~\bibnamefont {Blanch-Mercader}}, \bibinfo {author} {\bibfnamefont {G.}~\bibnamefont {Pernollet}}, \bibinfo {author} {\bibfnamefont {K.}~\bibnamefont {Kruse}},\ and\ \bibinfo {author} {\bibfnamefont {A.}~\bibnamefont {Roux}},\ }\href {https://doi.org/10.1038/s41563-022-01194-5} {\bibfield  {journal} {\bibinfo  {journal} {Nature Materials}\ }\textbf {\bibinfo {volume} {21}},\ \bibinfo {pages} {588} (\bibinfo {year} {2022})}\BibitemShut {NoStop}%
\bibitem [{\citenamefont {Blanch-Mercader}\ \emph {et~al.}(2021{\natexlab{a}})\citenamefont {Blanch-Mercader}, \citenamefont {Guillamat}, \citenamefont {Roux},\ and\ \citenamefont {Kruse}}]{Blanch-Mercader2021PRL}%
  \BibitemOpen
  \bibfield  {author} {\bibinfo {author} {\bibfnamefont {C.}~\bibnamefont {Blanch-Mercader}}, \bibinfo {author} {\bibfnamefont {P.}~\bibnamefont {Guillamat}}, \bibinfo {author} {\bibfnamefont {A.}~\bibnamefont {Roux}},\ and\ \bibinfo {author} {\bibfnamefont {K.}~\bibnamefont {Kruse}},\ }\href@noop {} {\bibfield  {journal} {\bibinfo  {journal} {Physical Review Letters}\ }\textbf {\bibinfo {volume} {126}},\ \bibinfo {pages} {028101} (\bibinfo {year} {2021}{\natexlab{a}})}\BibitemShut {NoStop}%
\bibitem [{\citenamefont {Blanch-Mercader}\ \emph {et~al.}(2021{\natexlab{b}})\citenamefont {Blanch-Mercader}, \citenamefont {Guillamat}, \citenamefont {Roux},\ and\ \citenamefont {Kruse}}]{Blanch-Mercader2021E}%
  \BibitemOpen
  \bibfield  {author} {\bibinfo {author} {\bibfnamefont {C.}~\bibnamefont {Blanch-Mercader}}, \bibinfo {author} {\bibfnamefont {P.}~\bibnamefont {Guillamat}}, \bibinfo {author} {\bibfnamefont {A.}~\bibnamefont {Roux}},\ and\ \bibinfo {author} {\bibfnamefont {K.}~\bibnamefont {Kruse}},\ }\href@noop {} {\bibfield  {journal} {\bibinfo  {journal} {Physical Review E}\ }\textbf {\bibinfo {volume} {103}},\ \bibinfo {pages} {012405} (\bibinfo {year} {2021}{\natexlab{b}})}\BibitemShut {NoStop}%
\bibitem [{\citenamefont {Endresen}\ \emph {et~al.}(2021)\citenamefont {Endresen}, \citenamefont {Kim}, \citenamefont {Pittman}, \citenamefont {Chen},\ and\ \citenamefont {Serra}}]{Endresen2021}%
  \BibitemOpen
  \bibfield  {author} {\bibinfo {author} {\bibfnamefont {K.~D.}\ \bibnamefont {Endresen}}, \bibinfo {author} {\bibfnamefont {M.~S.}\ \bibnamefont {Kim}}, \bibinfo {author} {\bibfnamefont {M.}~\bibnamefont {Pittman}}, \bibinfo {author} {\bibfnamefont {Y.}~\bibnamefont {Chen}},\ and\ \bibinfo {author} {\bibfnamefont {F.}~\bibnamefont {Serra}},\ }\href {https://doi.org/10.1039/d1sm00100k} {\bibfield  {journal} {\bibinfo  {journal} {Soft Matter}\ }\textbf {\bibinfo {volume} {17}},\ \bibinfo {pages} {5878} (\bibinfo {year} {2021})}\BibitemShut {NoStop}%
\bibitem [{\citenamefont {Kaiyrbekov}\ \emph {et~al.}(2023)\citenamefont {Kaiyrbekov}, \citenamefont {Endresen}, \citenamefont {Sullivan}, \citenamefont {Zheng}, \citenamefont {Chen}, \citenamefont {Serra},\ and\ \citenamefont {Camley}}]{Kaiyrbekov2023}%
  \BibitemOpen
  \bibfield  {author} {\bibinfo {author} {\bibfnamefont {K.}~\bibnamefont {Kaiyrbekov}}, \bibinfo {author} {\bibfnamefont {K.}~\bibnamefont {Endresen}}, \bibinfo {author} {\bibfnamefont {K.}~\bibnamefont {Sullivan}}, \bibinfo {author} {\bibfnamefont {Z.}~\bibnamefont {Zheng}}, \bibinfo {author} {\bibfnamefont {Y.}~\bibnamefont {Chen}}, \bibinfo {author} {\bibfnamefont {F.}~\bibnamefont {Serra}},\ and\ \bibinfo {author} {\bibfnamefont {B.~A.}\ \bibnamefont {Camley}},\ }\href {https://doi.org/10.1073/pnas.2301197120} {\bibfield  {journal} {\bibinfo  {journal} {Proceedings of the National Academy of Sciences of the United States of America}\ }\textbf {\bibinfo {volume} {120}},\ \bibinfo {pages} {1} (\bibinfo {year} {2023})}\BibitemShut {NoStop}%
\bibitem [{\citenamefont {Zhao}\ \emph {et~al.}(2024)\citenamefont {Zhao}, \citenamefont {Li}, \citenamefont {Yao}, \citenamefont {Zhao}, \citenamefont {Serra}, \citenamefont {Kawaguchi}, \citenamefont {Zhang}, \citenamefont {Chate},\ and\ \citenamefont {Sano}}]{Zihuiexper}%
  \BibitemOpen
  \bibfield  {author} {\bibinfo {author} {\bibfnamefont {Z.}~\bibnamefont {Zhao}}, \bibinfo {author} {\bibfnamefont {H.}~\bibnamefont {Li}}, \bibinfo {author} {\bibfnamefont {Y.}~\bibnamefont {Yao}}, \bibinfo {author} {\bibfnamefont {Y.}~\bibnamefont {Zhao}}, \bibinfo {author} {\bibfnamefont {F.}~\bibnamefont {Serra}}, \bibinfo {author} {\bibfnamefont {K.}~\bibnamefont {Kawaguchi}}, \bibinfo {author} {\bibfnamefont {H.}~\bibnamefont {Zhang}}, \bibinfo {author} {\bibfnamefont {H.}~\bibnamefont {Chate}},\ and\ \bibinfo {author} {\bibfnamefont {M.}~\bibnamefont {Sano}},\ }\href@noop {} {}\bibinfo {howpublished} {e-print \url{bioRxiv:10.1101/2024.08.28.610106}} (\bibinfo {year} {2024})\BibitemShut {NoStop}%
\bibitem [{\citenamefont {Ramaswamy}\ \emph {et~al.}(2003)\citenamefont {Ramaswamy}, \citenamefont {Simha},\ and\ \citenamefont {Toner}}]{ramaswamy2003active}%
  \BibitemOpen
  \bibfield  {author} {\bibinfo {author} {\bibfnamefont {S.}~\bibnamefont {Ramaswamy}}, \bibinfo {author} {\bibfnamefont {R.~A.}\ \bibnamefont {Simha}},\ and\ \bibinfo {author} {\bibfnamefont {J.}~\bibnamefont {Toner}},\ }\href@noop {} {\bibfield  {journal} {\bibinfo  {journal} {Europhysics Letters}\ }\textbf {\bibinfo {volume} {62}},\ \bibinfo {pages} {196} (\bibinfo {year} {2003})}\BibitemShut {NoStop}%
\bibitem [{\citenamefont {Marenduzzo}\ \emph {et~al.}(2007)\citenamefont {Marenduzzo}, \citenamefont {Orlandini}, \citenamefont {Cates},\ and\ \citenamefont {Yeomans}}]{marenduzzo2007steady}%
  \BibitemOpen
  \bibfield  {author} {\bibinfo {author} {\bibfnamefont {D.}~\bibnamefont {Marenduzzo}}, \bibinfo {author} {\bibfnamefont {E.}~\bibnamefont {Orlandini}}, \bibinfo {author} {\bibfnamefont {M.}~\bibnamefont {Cates}},\ and\ \bibinfo {author} {\bibfnamefont {J.}~\bibnamefont {Yeomans}},\ }\href@noop {} {\bibfield  {journal} {\bibinfo  {journal} {Physical Review E—Statistical, Nonlinear, and Soft Matter Physics}\ }\textbf {\bibinfo {volume} {76}},\ \bibinfo {pages} {031921} (\bibinfo {year} {2007})}\BibitemShut {NoStop}%
\bibitem [{\citenamefont {Giomi}\ \emph {et~al.}(2011)\citenamefont {Giomi}, \citenamefont {Mahadevan}, \citenamefont {Chakraborty},\ and\ \citenamefont {Hagan}}]{giomi2011excitable}%
  \BibitemOpen
  \bibfield  {author} {\bibinfo {author} {\bibfnamefont {L.}~\bibnamefont {Giomi}}, \bibinfo {author} {\bibfnamefont {L.}~\bibnamefont {Mahadevan}}, \bibinfo {author} {\bibfnamefont {B.}~\bibnamefont {Chakraborty}},\ and\ \bibinfo {author} {\bibfnamefont {M.~F.}\ \bibnamefont {Hagan}},\ }\href@noop {} {\bibfield  {journal} {\bibinfo  {journal} {Physical review letters}\ }\textbf {\bibinfo {volume} {106}},\ \bibinfo {pages} {218101} (\bibinfo {year} {2011})}\BibitemShut {NoStop}%
\bibitem [{\citenamefont {Giomi}\ \emph {et~al.}(2013)\citenamefont {Giomi}, \citenamefont {Bowick}, \citenamefont {Ma},\ and\ \citenamefont {Marchetti}}]{giomi2013defect}%
  \BibitemOpen
  \bibfield  {author} {\bibinfo {author} {\bibfnamefont {L.}~\bibnamefont {Giomi}}, \bibinfo {author} {\bibfnamefont {M.~J.}\ \bibnamefont {Bowick}}, \bibinfo {author} {\bibfnamefont {X.}~\bibnamefont {Ma}},\ and\ \bibinfo {author} {\bibfnamefont {M.~C.}\ \bibnamefont {Marchetti}},\ }\href@noop {} {\bibfield  {journal} {\bibinfo  {journal} {Physical review letters}\ }\textbf {\bibinfo {volume} {110}},\ \bibinfo {pages} {228101} (\bibinfo {year} {2013})}\BibitemShut {NoStop}%
\bibitem [{\citenamefont {Shi}\ and\ \citenamefont {Ma}(2013)}]{Shi2013}%
  \BibitemOpen
  \bibfield  {author} {\bibinfo {author} {\bibfnamefont {X.~Q.}\ \bibnamefont {Shi}}\ and\ \bibinfo {author} {\bibfnamefont {Y.~Q.}\ \bibnamefont {Ma}},\ }\href {https://doi.org/10.1038/ncomms4013} {\bibfield  {journal} {\bibinfo  {journal} {Nature Communications}\ }\textbf {\bibinfo {volume} {4}},\ \bibinfo {pages} {1} (\bibinfo {year} {2013})}\BibitemShut {NoStop}%
\bibitem [{\citenamefont {Gao}\ \emph {et~al.}(2015)\citenamefont {Gao}, \citenamefont {Blackwell}, \citenamefont {Glaser}, \citenamefont {Betterton},\ and\ \citenamefont {Shelley}}]{gao2015multiscale}%
  \BibitemOpen
  \bibfield  {author} {\bibinfo {author} {\bibfnamefont {T.}~\bibnamefont {Gao}}, \bibinfo {author} {\bibfnamefont {R.}~\bibnamefont {Blackwell}}, \bibinfo {author} {\bibfnamefont {M.~A.}\ \bibnamefont {Glaser}}, \bibinfo {author} {\bibfnamefont {M.~D.}\ \bibnamefont {Betterton}},\ and\ \bibinfo {author} {\bibfnamefont {M.~J.}\ \bibnamefont {Shelley}},\ }\href@noop {} {\bibfield  {journal} {\bibinfo  {journal} {Physical review letters}\ }\textbf {\bibinfo {volume} {114}},\ \bibinfo {pages} {048101} (\bibinfo {year} {2015})}\BibitemShut {NoStop}%
\bibitem [{\citenamefont {Giomi}(2015)}]{giomi2015geometry}%
  \BibitemOpen
  \bibfield  {author} {\bibinfo {author} {\bibfnamefont {L.}~\bibnamefont {Giomi}},\ }\href@noop {} {\bibfield  {journal} {\bibinfo  {journal} {Physical Review X}\ }\textbf {\bibinfo {volume} {5}},\ \bibinfo {pages} {031003} (\bibinfo {year} {2015})}\BibitemShut {NoStop}%
\bibitem [{\citenamefont {Green}\ \emph {et~al.}(2017)\citenamefont {Green}, \citenamefont {Toner},\ and\ \citenamefont {Vitelli}}]{Green2017}%
  \BibitemOpen
  \bibfield  {author} {\bibinfo {author} {\bibfnamefont {R.}~\bibnamefont {Green}}, \bibinfo {author} {\bibfnamefont {J.}~\bibnamefont {Toner}},\ and\ \bibinfo {author} {\bibfnamefont {V.}~\bibnamefont {Vitelli}},\ }\href {https://doi.org/10.1103/PhysRevFluids.2.104201} {\bibfield  {journal} {\bibinfo  {journal} {Physical Review Fluids}\ }\textbf {\bibinfo {volume} {2}},\ \bibinfo {pages} {20} (\bibinfo {year} {2017})}\BibitemShut {NoStop}%
\bibitem [{\citenamefont {Maitra}\ \emph {et~al.}(2018)\citenamefont {Maitra}, \citenamefont {Srivastava}, \citenamefont {{Cristina Marchetti}}, \citenamefont {Lintuvuori}, \citenamefont {Ramaswamy},\ and\ \citenamefont {Lenz}}]{Maitra2018}%
  \BibitemOpen
  \bibfield  {author} {\bibinfo {author} {\bibfnamefont {A.}~\bibnamefont {Maitra}}, \bibinfo {author} {\bibfnamefont {P.}~\bibnamefont {Srivastava}}, \bibinfo {author} {\bibfnamefont {M.}~\bibnamefont {{Cristina Marchetti}}}, \bibinfo {author} {\bibfnamefont {J.~S.}\ \bibnamefont {Lintuvuori}}, \bibinfo {author} {\bibfnamefont {S.}~\bibnamefont {Ramaswamy}},\ and\ \bibinfo {author} {\bibfnamefont {M.}~\bibnamefont {Lenz}},\ }\href {https://doi.org/10.1073/pnas.1720607115} {\bibfield  {journal} {\bibinfo  {journal} {Proceedings of the National Academy of Sciences of the United States of America}\ }\textbf {\bibinfo {volume} {115}},\ \bibinfo {pages} {6934} (\bibinfo {year} {2018})}\BibitemShut {NoStop}%
\bibitem [{\citenamefont {Doostmohammadi}\ \emph {et~al.}(2018)\citenamefont {Doostmohammadi}, \citenamefont {Ign{\'e}s-Mullol}, \citenamefont {Yeomans},\ and\ \citenamefont {Sagu{\'e}s}}]{doostmohammadi2018}%
  \BibitemOpen
  \bibfield  {author} {\bibinfo {author} {\bibfnamefont {A.}~\bibnamefont {Doostmohammadi}}, \bibinfo {author} {\bibfnamefont {J.}~\bibnamefont {Ign{\'e}s-Mullol}}, \bibinfo {author} {\bibfnamefont {J.~M.}\ \bibnamefont {Yeomans}},\ and\ \bibinfo {author} {\bibfnamefont {F.}~\bibnamefont {Sagu{\'e}s}},\ }\href@noop {} {\bibfield  {journal} {\bibinfo  {journal} {Nature communications}\ }\textbf {\bibinfo {volume} {9}},\ \bibinfo {pages} {3246} (\bibinfo {year} {2018})}\BibitemShut {NoStop}%
\bibitem [{\citenamefont {Balasubramaniam}\ \emph {et~al.}(2021)\citenamefont {Balasubramaniam}, \citenamefont {Doostmohammadi}, \citenamefont {Saw}, \citenamefont {Narayana}, \citenamefont {Mueller}, \citenamefont {Dang}, \citenamefont {Thomas}, \citenamefont {Gupta}, \citenamefont {Sonam}, \citenamefont {Yap} \emph {et~al.}}]{balasubramaniam2021investigating}%
  \BibitemOpen
  \bibfield  {author} {\bibinfo {author} {\bibfnamefont {L.}~\bibnamefont {Balasubramaniam}}, \bibinfo {author} {\bibfnamefont {A.}~\bibnamefont {Doostmohammadi}}, \bibinfo {author} {\bibfnamefont {T.~B.}\ \bibnamefont {Saw}}, \bibinfo {author} {\bibfnamefont {G.~H. N.~S.}\ \bibnamefont {Narayana}}, \bibinfo {author} {\bibfnamefont {R.}~\bibnamefont {Mueller}}, \bibinfo {author} {\bibfnamefont {T.}~\bibnamefont {Dang}}, \bibinfo {author} {\bibfnamefont {M.}~\bibnamefont {Thomas}}, \bibinfo {author} {\bibfnamefont {S.}~\bibnamefont {Gupta}}, \bibinfo {author} {\bibfnamefont {S.}~\bibnamefont {Sonam}}, \bibinfo {author} {\bibfnamefont {A.~S.}\ \bibnamefont {Yap}}, \emph {et~al.},\ }\href@noop {} {\bibfield  {journal} {\bibinfo  {journal} {Nature Materials}\ }\textbf {\bibinfo {volume} {20}},\ \bibinfo {pages} {1156} (\bibinfo {year} {2021})}\BibitemShut {NoStop}%
\bibitem [{\citenamefont {Simha}\ and\ \citenamefont {Ramaswamy}(2002)}]{simha2002}%
  \BibitemOpen
  \bibfield  {author} {\bibinfo {author} {\bibfnamefont {R.~A.}\ \bibnamefont {Simha}}\ and\ \bibinfo {author} {\bibfnamefont {S.}~\bibnamefont {Ramaswamy}},\ }\href@noop {} {\bibfield  {journal} {\bibinfo  {journal} {Physical review letters}\ }\textbf {\bibinfo {volume} {89}},\ \bibinfo {pages} {058101} (\bibinfo {year} {2002})}\BibitemShut {NoStop}%
\bibitem [{\citenamefont {Vafa}\ \emph {et~al.}(2021)\citenamefont {Vafa}, \citenamefont {Bowick}, \citenamefont {Shraiman},\ and\ \citenamefont {Marchetti}}]{Vafa2021}%
  \BibitemOpen
  \bibfield  {author} {\bibinfo {author} {\bibfnamefont {F.}~\bibnamefont {Vafa}}, \bibinfo {author} {\bibfnamefont {M.~J.}\ \bibnamefont {Bowick}}, \bibinfo {author} {\bibfnamefont {B.~I.}\ \bibnamefont {Shraiman}},\ and\ \bibinfo {author} {\bibfnamefont {M.~C.}\ \bibnamefont {Marchetti}},\ }\href {https://doi.org/10.1039/d0sm02027c} {\bibfield  {journal} {\bibinfo  {journal} {Soft Matter}\ }\textbf {\bibinfo {volume} {17}},\ \bibinfo {pages} {3068} (\bibinfo {year} {2021})}\BibitemShut {NoStop}%
\bibitem [{\citenamefont {Killeen}\ \emph {et~al.}(2022)\citenamefont {Killeen}, \citenamefont {Bertrand},\ and\ \citenamefont {Lee}}]{Killeen2022}%
  \BibitemOpen
  \bibfield  {author} {\bibinfo {author} {\bibfnamefont {A.}~\bibnamefont {Killeen}}, \bibinfo {author} {\bibfnamefont {T.}~\bibnamefont {Bertrand}},\ and\ \bibinfo {author} {\bibfnamefont {C.~F.}\ \bibnamefont {Lee}},\ }\href {https://doi.org/10.1103/PhysRevLett.128.078001} {\bibfield  {journal} {\bibinfo  {journal} {Physical Review Letters}\ }\textbf {\bibinfo {volume} {128}},\ \bibinfo {pages} {78001} (\bibinfo {year} {2022})}\BibitemShut {NoStop}%
\bibitem [{\citenamefont {Zhang}\ and\ \citenamefont {Yeomans}(2023)}]{Zhang2023}%
  \BibitemOpen
  \bibfield  {author} {\bibinfo {author} {\bibfnamefont {G.}~\bibnamefont {Zhang}}\ and\ \bibinfo {author} {\bibfnamefont {J.~M.}\ \bibnamefont {Yeomans}},\ }\href {https://doi.org/10.1103/PhysRevLett.130.038202} {\bibfield  {journal} {\bibinfo  {journal} {Physical Review Letters}\ }\textbf {\bibinfo {volume} {130}},\ \bibinfo {pages} {38202} (\bibinfo {year} {2023})}\BibitemShut {NoStop}%
\bibitem [{\citenamefont {Patelli}\ \emph {et~al.}(2019)\citenamefont {Patelli}, \citenamefont {Djafer-Cherif}, \citenamefont {Aranson}, \citenamefont {Bertin},\ and\ \citenamefont {Chat{\'{e}}}}]{Patelli2019}%
  \BibitemOpen
  \bibfield  {author} {\bibinfo {author} {\bibfnamefont {A.}~\bibnamefont {Patelli}}, \bibinfo {author} {\bibfnamefont {I.}~\bibnamefont {Djafer-Cherif}}, \bibinfo {author} {\bibfnamefont {I.~S.}\ \bibnamefont {Aranson}}, \bibinfo {author} {\bibfnamefont {E.}~\bibnamefont {Bertin}},\ and\ \bibinfo {author} {\bibfnamefont {H.}~\bibnamefont {Chat{\'{e}}}},\ }\href {https://doi.org/10.1103/PhysRevLett.123.258001} {\bibfield  {journal} {\bibinfo  {journal} {Physical Review Letters}\ }\textbf {\bibinfo {volume} {123}},\ \bibinfo {pages} {1} (\bibinfo {year} {2019})}\BibitemShut {NoStop}%
\bibitem [{SUP()}]{SUPP}%
  \BibitemOpen
  \href@noop {} {}\bibinfo {note} {See Supplemental Material at URL,}\BibitemShut {NoStop}%
\bibitem [{\citenamefont {De~Gennes}\ and\ \citenamefont {Prost}(1993)}]{deGennes1993}%
  \BibitemOpen
  \bibfield  {author} {\bibinfo {author} {\bibfnamefont {P.-G.}\ \bibnamefont {De~Gennes}}\ and\ \bibinfo {author} {\bibfnamefont {J.}~\bibnamefont {Prost}},\ }\href@noop {} {\emph {\bibinfo {title} {The physics of liquid crystals}}},\ \bibinfo {number} {83}\ (\bibinfo  {publisher} {Oxford university press},\ \bibinfo {year} {1993})\BibitemShut {NoStop}%
\bibitem [{\citenamefont {Beris}\ and\ \citenamefont {Edwards}(1994)}]{beris1994thermodynamics}%
  \BibitemOpen
  \bibfield  {author} {\bibinfo {author} {\bibfnamefont {A.~N.}\ \bibnamefont {Beris}}\ and\ \bibinfo {author} {\bibfnamefont {B.~J.}\ \bibnamefont {Edwards}},\ }\href@noop {} {\emph {\bibinfo {title} {Thermodynamics of flowing systems: with internal microstructure}}},\ \bibinfo {number} {36}\ (\bibinfo  {publisher} {Oxford University Press, USA},\ \bibinfo {year} {1994})\BibitemShut {NoStop}%
\bibitem [{Note1()}]{Note1}%
  \BibitemOpen
  \bibinfo {note} {One quantitative difference is that a smaller $C_p$ value is sufficient, for the simplified equations, to fully impose the external pattern and in particular to avoid that the $+1$ defect splits into two $+\protect \tfrac {1}{2}$ ones.}\BibitemShut {Stop}%
\end{thebibliography}%

\onecolumngrid

\vspace{12pt}
\noindent\hrulefill \hspace{24pt} {\bf End Matter} \hspace{24pt} \hrulefill
\vspace{14pt}

\begin{figure*}[h!]   
\includegraphics[width=\textwidth]{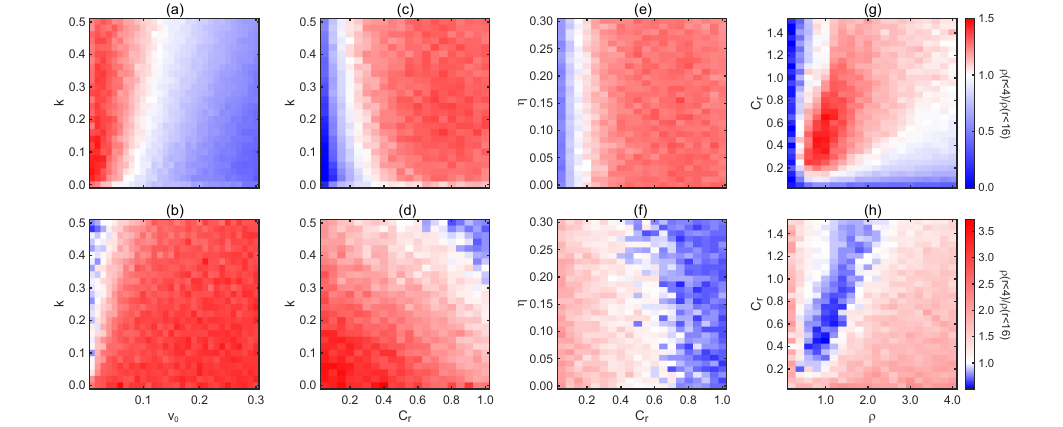}
\caption{\label{figEM1} Phase diagrams of the particle model in various parameter planes (same conditions as for Fig.~\ref{fig2}(d-f)). Upper/low row: target/aster field pattern. Accumulation (red) and depletion (blue) are evaluated by the averaged density ratio between the core region ($r<4$) and whole region ($r<16$). 
}
\end{figure*}

\twocolumngrid

{\it Appendix A: Complementary results at particle level.---}  Phase diagrams similar those presented in Fig.~\ref{fig2}(d-f), but in various other parameter planes are presented in Fig.~\ref{figEM1}. For each parameter plane shown, the top panel is for the target pattern, with the bottom panel for the aster pattern.

\begin{figure*}[th]
\includegraphics[width=\textwidth]{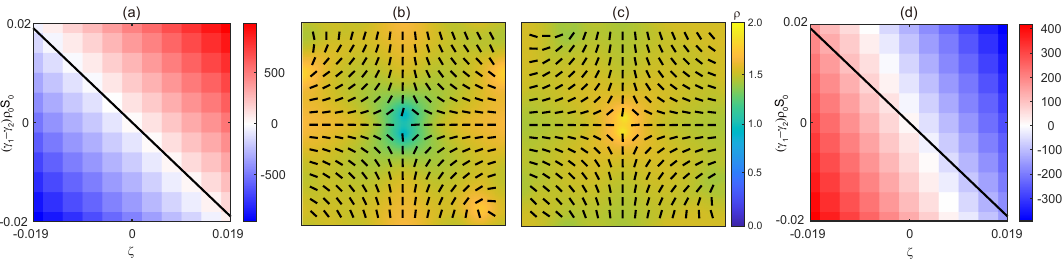}
\caption{\label{figEM2} 
Further simulations of Eqs.~(\ref{eqrho2}-\ref{eqQ}) (similar to those in Fig.~\ref{fig3}(a-c)).
(a): Phase diagram for aster pattern in the $(\zeta, \gamma_1-\gamma_2)$ plane. (same conditions as for Fig.~\ref{fig3}(a)($v_0\sim0.25$) except for $\zeta, \gamma_1-\gamma_2$). 
(b): Density and orientation field of $-1$ defect exhibiting depletion
($\zeta=0, \gamma_1=0.3, \gamma_2=-0.1$).
(c): Density and orientation field of $-1$ defect exhibiting accumulation
($\zeta=0, \gamma_1=-0.3, \gamma_2=-0.1$).
Note that the density distribution is not axisymmetric. 
(d): Phase diagram for the $-1$ defect in $(\zeta, \gamma_1-\gamma_2)$ plane. 
In (a) and (d), accumulation (red) and depletion (blue) are evaluated by the excess or missing mass $\rho -\rho_0$ integrated inside a circular region ($r<16$). The solid line is the prediction of the transition between accumulation and prediction. Simulation parameters for (b-d): all coefficients in Eqs.~(\ref{eqrho2}-\ref{eqQ}) are calculated from the particle-level 
$\rho=1.5, a=0.4, v_0=0.5, b_1=0.25, L=32$, except $\nu=0.25$ and $\zeta, \gamma_1, \gamma_2$ given above.}
\end{figure*}

{\it Appendix B: Complementary information about continuous theory.---}
Here we give some important points in the derivation of the continuous theory (\ref{eqrho}-\ref{eqf2}) 
that is detailed in \cite{SUPP} and follows  \cite{Patelli2019}.
In the Boltzmann equation \eqref{eqB}, the collision integral reads:
\begin{eqnarray}
	\label{eq:Collision_Integral-real}
&& I_{\rm col}[f] = \!\! \int_{-\pi}^{\pi}  \!\!\!\! d\theta_1 \!\! \int_{-\pi}^{\pi} \!\!\!\! d\theta_2 f(\mathbf{r},\theta_1) \!\! \int_0^{\infty} \!\!\!\! ds\, s 
	\!\! \int_{-\pi}^{\pi} \!\!\!\!d\phi \, K(s,\phi,\theta_1,\theta_2) \nonumber \\
&& \;\; \times f(\mathbf{r}\!+\!s\mathbf{e}(\phi),\theta_2) 
[\langle \hat{\delta}\big(\Psi(\theta_1,\theta_2)\!+\!\sigma \!-\!\theta\big)\rangle_{\psi} \!-\! 
\hat{\delta}(\theta_1\!-\!\theta)] 
\end{eqnarray}
where $\Psi$ is the $\pi$-periodic nematic alignment function
($\Psi(\theta_1,\theta_2)=\frac{1}{2}(\theta_2-\theta_1)$ for $-\frac{\pi}{2} < \theta_2-\theta_1< \frac{\pi}{2}$),
and $K$ is the kernel function
\begin{equation}
    K = 2g(s) |\sin \frac{\theta_1 -\theta_2}{2}|\cos(\phi - \theta_{12})\Theta[\cos(\phi-\theta_{12})].
\end{equation}
which codes for the repulsive torques between particles, 
 depends on the relative position $\mathbf{r'}\!-\!\mathbf{r} \equiv s\mathbf{e}(\phi)$ of the two colliding particles,  
and incorporates the distance-dependent weighting function $g(s)$.

Upon expanding $f(\mathbf{r}\!+\!s\mathbf{e}(\phi),\theta_2)$ at first order in $s$, 
and Fourier transforming \eqref{eqB} into a hierarchy for the fields $f_k$, 
$g(s)$ appears only via the first two  of its moments $b_{n} = \int_{0}^{s} s^{n+1} g(s) ds$:
$b_0$ that is set to unity in all generality, and $b_1$ that is thus the only microscopic-level coefficient
controlling the strength of repulsive torques, akin to $C_r$ in the particle-level model.

The coefficients appearing in Eqs.~(\ref{eqrho}-\ref{eqf2}) depend on the particle-level 
parameters $v_0$, $a$, $b_1$, $C_p$, and the modes $P_k$ of the noise distribution $\psi$ 
($P_k = \int_{\infty}^{\infty} d\psi P(\psi) \exp ik\psi$), which we take to be a Gaussian of rms $\eta$ in numerical 
applications. Their exact expressions are given in \cite{SUPP}, but here we give the most important ones:
\begin{eqnarray}
\zeta[\rho] &=& \frac{v_0}{2} - \frac{b_1}{6}\sqrt{2} P_1 \rho, \nonumber \\
\gamma_1 &=& \frac{4}{3\pi}(P_1 -\frac{2}{7})\frac{v_0 - \sqrt{2}b_1 P_3 \rho_0}{\rho_0 \frac{272}{35\pi}-P_3 + 1 +2a} + \frac{b_1}{6}\sqrt{2} P_1,  \nonumber \\
\gamma_2 &=& -\frac{b_1}{10}\sqrt{2} P_1 <0,  \nonumber   
\end{eqnarray}

{\it Appendix C:  Accumulation-depletion by a $-1$ defect.---} 
Departing from the coefficients dictated by our particle-level model, 
we independently varied $\zeta$ and $\gamma_1-\gamma_2$ in the generalized functional space.
This allowed us to explore parameter regions that are not easily accessible within our self-imposed particle-level constraints.

As shown in Fig.~\ref{fig3}(c), simulations of Eqs.~(\ref{eqrho2}-\ref{eqQ})
hardly show density depletion at the core region of the aster pattern because the coefficient $\gamma_1 - \gamma_2$ is mostly positive and rarely takes large negative values (when varying the parameters of the particle-level model).
The phase diagram in the $(\zeta, \gamma_1 - \gamma_2)$ plane, on the other hand,
shows a large region of depletion when these parameters are varied independently, and
the prediction by Eq.~\eqref{eqfr} is slightly below the actual transition from accumulation to depletion in aster 
(Fig.~\ref{figEM2}(a)).
The behavior near a $-1$ defect was also investigated this way. 
We observe mostly depletion near its core with the particle-level-dictated parameters, 
but when $\gamma_1 -\gamma_2$ is varied over a wider range, accumulation is 
easily observed and the transition line deviates from the prediction (Fig.~\ref{figEM2}(d). 
We attribute this discrepancy to the assumption 
of axisymmetric solutions used in obtaining \eqref{eqfr}, which is of course broken near a $-1$ defect.

\end{document}


\title{Supplemental Material for \\ ``Integer Topological Defects Reveal Anti-Symmetric Forces in Active Nematics''}

\author{Zihui Zhao*}
\affiliation{School of Physics and Astronomy, Institute of Natural Sciences, Shanghai Jiao Tong University, Shanghai 200240, China}
\altaffiliation[]{contributed equally}

\author{Yisong Yao*}
\affiliation{School of Physics and Astronomy, Institute of Natural Sciences, Shanghai Jiao Tong University, Shanghai 200240, China}
\altaffiliation[]{contributed equally}

\author{He Li}
\affiliation{School of Physics and Astronomy, Institute of Natural Sciences, Shanghai Jiao Tong University, Shanghai 200240, China}
\affiliation{Institut de Génétique et de Biologie Moléculaire et Cellulaire, Illkirch, France}

\author{Yongfeng Zhao}
\affiliation{School of Physics and Astronomy, Institute of Natural Sciences, Shanghai Jiao Tong University, Shanghai 200240, China}
\affiliation{Center for Soft Condensed Matter Physics \& Interdisciplinary Research, Soochow University, Suzhou 215006, China}

\author{Yujia~Wang}
\affiliation{Center for Soft Condensed Matter Physics \& Interdisciplinary Research, Soochow University, Suzhou 215006, China}

\author{Hepeng Zhang}
\affiliation{School of Physics and Astronomy, Institute of Natural Sciences, Shanghai Jiao Tong University, Shanghai 200240, China}

\author{Hugues Chat\'{e}}
\affiliation{Service de Physique de l'Etat Condens\'e, CEA, CNRS Universit\'e Paris-Saclay, CEA-Saclay, 91191 Gif-sur-Yvette, France}
\affiliation{Computational Science Research Center, Beijing 100094, China}

\author{Masaki Sano}
\affiliation{School of Physics and Astronomy, Institute of Natural Sciences, Shanghai Jiao Tong University, Shanghai 200240, China}
\affiliation{Universal Biology Institute, The University of Tokyo, Bunkyo-ku, Tokyo 113-0033, Japan}

\date{\today}

\maketitle

\section{Integer defects in standard active nematic theory}

When cells in monolayers have head-tail symmetry and orientational order, active matter theory often relies on classic continuum nematohydrodynamics complemented by an active force (or stress)  $\bm{f}^a$,
defined by the spatial derivative of the nematic tensor order parameter $\bm{Q}$ as, $f^a_i = - \zeta \nabla_j Q_{ij}$, with
\begin{equation}
\label{eq:S1}
\bm{Q} = \frac{d}{d-1}S(\bm{n} \bm{n} - \frac{\bm{I}}{d})
\end{equation}
where $d$ is the spatial dimension, $\bm{n}$ is the director ($\bm{n} = -\bm{n}$) expressing local orientation, $S$ is
the scalar amplitude of the nematic order. 
For +1 defects in 2D, the local orientation around the origin is given by 
$\theta = \phi + \theta_0$ with the polar angle $\phi$ and the tilt angle $\theta_0$. Substituting 
$\bm{n} = (\cos \theta, \sin \theta)^t$ into Eq.\eqref{eq:S1} leads
\begin{equation}
\bm{Q}^{+1} = S(r)
\begin{pmatrix}
   \cos 2(\phi + \theta_0) & \sin 2(\phi + \theta_0) \\
   \sin 2(\phi + \theta_0) & -\cos 2(\phi + \theta_0)
\end{pmatrix}.
\end{equation}
Therefore, the active forces around +1 defects are calculated as
\begin{equation}
\bm{f}^a = -\zeta \nabla \cdot \bm{Q} = -\zeta [S'(r) + \frac{2S(r)}{r}](\cos 2\theta_0 \hat{\bm e}_r + \sin 2\theta_0 \hat{\bm e}_{\phi}),
\end{equation}
where $S' = \frac{dS}{dr}$, $\hat{\bm e}_r$ and $\hat{\bm e}_{\phi}$ are the unit vectors for radial and azimuthal direction, respectively.

\section{Derivation of continuous theory}
\subsection{Boltzmann equation with external field}

The Boltzmann equation under the external field $\bm{F}$ is given in general by
\begin{equation}
\partial_t f + \bm{v} \cdot \frac{\partial f}{\partial \bm{x}} + \bm{F} \cdot \frac{\partial f}{\partial \bm{p}} = I_{col}[f] ,
\end{equation}
where  $\bm{v}  = \partial_t \bm{x}$ and $\bm{F}  = \partial_t \bm{p}$.
However, when the angle dynamics of the particles are introduced through the interaction between particles or with obstacles, stochastic dynamics of the angle change need to be considered. 
If the effect of the external field
is to induce nematic alignment to the locally defined angle $\theta_p(\bm{r})$. Stochastic dynamics of the angle $\theta$ of each particle is given by
\begin{equation}
\partial_t \theta = \omega(\theta) + \eta = C_p \sin 2(\theta_p - \theta) + \eta,
\end{equation}
here the angular speed $\omega(\theta)$ is assumed to obey $C_p \sin 2(\theta_p - \theta)$ with $C_p$ the strength of the external field and $\eta$ is Gaussian white noise. Angle change due to interaction with other particles is included in the collision term.
In such a case, the streaming part of the Boltzmann equation needs to consider stochastic dynamics on the angle change. In general, velocity and angle changes in a small time step are expressed by:
\begin{equation}
f(\bm{r}', \theta', t \!+\! \Delta t) \!=\!\! \int_{0}^{\infty}\!\!\!\! dv\!\! \int_{-\pi}^{\pi} \!\!\!\!d\theta\, \Phi(\bm{r}' \!-\! \bm{r}, \theta' \!-\!\theta, \Delta t) f(\bm{r}, \theta, t) ,
\end{equation}
where $\bm{r}' = \bm{r} + \bm{v}\Delta t$, $\theta' = \theta + \omega(\theta) \Delta t$, $\Phi$ is the transition probability for the position and angle dynamics.
Employing the Taylor expansion of $\Phi$ for a small $\Delta t$, 
\begin{eqnarray}
\Phi(\bm{r}' - \bm{r}, \theta' -\theta, \Delta t) &=& 1 + \partial_{\bm{r}} \Phi(\bm{r}, \theta,\Delta t) \cdot \bm{v}\Delta t  \nonumber \\
&+& \partial_{\theta}  \Phi(\bm{r}, \theta,\Delta t) \omega(\theta) \Delta t .
\end{eqnarray}
The time derivative of $f$ leads to
\begin{equation}
\partial_t f \!\!=\!\! \int_{0}^{\infty} \!\!\!\!dv\!\! \int_{-\pi}^{\pi} \!\!\!\!d\theta\, [ \partial_{\bm{r}} \Phi(\bm{r}, \theta,\Delta t) \cdot \bm{v} + \partial_{\theta}  \Phi(\bm{r}, \theta,\Delta t) \omega(\theta) ] f.  \label{eqdtf}
\end{equation}
Here we employ the delta function for $\Phi$.
\begin{equation}
\Phi(\bm{r}' - \bm{r}, \theta' -\theta, \Delta t) = \delta(\bm{r}' - \bm{r} -\bm{v}_0\Delta t) \delta(\theta' -\theta -\omega(\theta)\Delta t).
\end{equation}
With using a property of delta  function, $\int_{-\infty}^{\infty} g(x) \delta'(x - x_0) dx = - g'(x_0)$, Eq.(\ref{eqdtf}) becomes
\begin{equation}
\partial_t f = - \bm{v}_0 \cdot \nabla_{\bm{r}} f - \partial_{\theta} [\omega(\theta) f] .
\end{equation}
Therefore, the Boltzmann equation under the external field is given as
\begin{eqnarray}
&&\partial_t f + v_0 \bm{e}(\theta) \cdot \nabla f + \partial_{\theta} [C_p \sin 2(\theta_p - \theta) f ]  \nonumber \\
&=& I_{col}[f] + \lambda_0 \bigl [\langle f(\theta - \psi)\rangle_{\psi} - f(\psi) \bigr] \nonumber \\
&+& a \bigl [f(\theta + \pi) - f(\theta) \bigr].
\end{eqnarray}
The collision integral is
\begin{align}
I_{col}[f]=\int_{-\pi}^{\pi} d \theta_1 \int_{-\pi}^{\pi}d\theta_2 f(\bm{r},\theta_1) \int_{0}^{\infty} ds \,s \nonumber \\
\times \int_{-\pi}^{\pi} K(s, \phi, \theta_1, \theta_2)f(\bm{r}+s\bm{e}(\phi),\theta_2) \nonumber \\
\times [\langle\delta(\Psi(\theta_1,\theta_2)+\psi-\theta)\rangle_{\psi} - \delta(\theta_1 - \theta)],
\end{align}
where $\Psi$ is defined by
$\Psi(\theta_1,\theta_2)= \frac{1}{2}(\theta_2 - \theta_1)$ for ($\-\pi/2 < \theta_2 -\theta_1 < \pi/2$), and the collision kernel,
\begin{equation}
    K = 2g(s) |\sin \frac{\theta_1 -\theta_2}{2}|\cos(\phi - \theta_{12})\Theta[\cos(\phi-\theta_{12})].
\end{equation}

By adopting a Fourier expansion,
$f(\bm{r}, \theta, t) =$ ${\displaystyle \frac{1}{2\pi}\sum_{k=-\infty}^{\infty} f_k(\bm{r},t) e^{-ik\theta}}$ and $f_k(\bm{r}, t) = \int_{-\pi}^{\pi} f(\bm{r}, \theta, t)  e^{ik\theta} d\theta$, we obtain
\begin{eqnarray}
    \partial_t f_k &+& \frac{v_0}{2}(\nabla f_{k-1} + \nabla^* f_{k+1}) -  \frac{k C_p}{2} (e^{2i \theta_p} f_{k-2} - e^{-2i \theta_p} f_{k+2})  \nonumber \\
    &=& (P_k -1 - a|1 - (-1)^k|) f_k + \int_{-\pi}^{\pi} I_{col}[f]  e^{ik\theta} d\theta
\end{eqnarray}
here we set $\lambda_0=1$ and $P_k = \int_{\infty}^{\infty} d\psi P(\psi) \exp ik\psi$ are the Fourier modes of noise distributions . For Gaussian noise,
\begin{equation}
P_1 = e^{-\frac{\eta^2}{2}}; \quad P_2 =  e^{-\frac{(2\eta)^2}{2}}; \quad P_3 = e^{-\frac{(3\eta)^2}{2}}. \nonumber
\end{equation}

 Fourier components are related to density, $f_0 = \rho$, the polarity, $\bm{w} \equiv \frac{\rho \bm{u}}{u_0}= (\mathcal{R}(f_1), \mathcal{I}(f_1))$, and the nematic order tensor, $\tilde{Q}_{xx} \equiv \rho Q_{xx} = \mathcal{R}(f_2)$ and $\tilde{Q}_{xy} \equiv \rho Q_{xy} = \mathcal{I}(f_2)$.

The collision integral was evaluated following Patelli {\it et al.}~\cite{Patelli2019} where
$b_n$ was defined as
\begin{equation}
b_{n} = \int_{0}^{s} s^{n+1} g(s) ds,
\end{equation}
where $g(s)$ is the integrable function modeling soft repulsion in the interval $[0, d_0]$,
and we set $b_0=1$.

After some calculations, we obtain
\begin{eqnarray}
\partial_t f_k &=& -\frac{v_0}{2} (\nabla f_{k-1} + \nabla^* f_{k-1}) +(P_k -1 - a|1 - (-1)^k|) f_k  \nonumber \\
&+& b_0 \sum_{q} J_{k,q,0} f_{k-q} f_q + \frac{b_1}{2} \sum_{q} (J_{k,q,1}f_{k-q+1} \nabla^* f_q \nonumber \\
&+& J_{k,q,-1} f_{k-q-1} \nabla f_q) 
+ \frac{k C_p}{2} (e^{2i \theta_p} f_{k-2} - e^{-2i \theta_p} f_{k+2}) . \nonumber
\end{eqnarray}

Here the notation of the coefficients are
\begin{equation}
J_{k,q,n} = \frac{1}{\pi} A_n e^{-i n\pi/2}\int_{0}^{2\pi} d\Delta \sin \frac{\Delta}{2} [P_k e^{ik \mathit{H}(\Delta)} -1 ]e^{-iq\Delta} e^{i\frac{n}{2}\Delta}
\end{equation}
with
\begin{equation}
  \begin{cases}
    A_n = \frac{2}{1 - n^2}\cos \frac{n\pi}{2}     & \text{if $|n| \ne 1$} \\
    A_n = \frac{\pi}{2}                                   & \text{if $|n| = 1$} 
  \end{cases}
  \end{equation}
   
\begin{equation}
        \centering
  \mathit{H}(\Delta) = \frac{\Delta}{2} -
  \begin{cases}
    0,               & \text{$0<\Delta<\frac{\pi}{2}$} \\
    \frac{\pi}{2},  & \text{$\frac{\pi}{2}<\Delta<\frac{3\pi}{2}$} \\
    \pi .            & \text{$\frac{3\pi}{2}<\Delta<2\pi$}
  \end{cases}
\end{equation}

With the scaling anzatz, $\nabla \sim \partial_t \sim \delta \rho \sim \varepsilon, f_{2k-1} \sim f_{2k} \sim \varepsilon^{|k|}, C_p \sim \varepsilon^2$, 
we obtain the truncated hierarchy at the $\mathit{Q}(\varepsilon^3)$ order,
\begin{eqnarray}
\partial_t \rho &=& -\frac{v_0}{2} (\nabla f^*_1 + \nabla^* f_1 ), \\
\partial_t f_1 &=& \mu_1[\rho] f_1 - \frac{v_0}{2} (\nabla \rho + \nabla^* f_2 ) \nonumber \\
&+& b_0 (C_{1,2,0}f^*_1 f_2 + C_{1,3,0} f_3 f^*_2) \nonumber \\
&+& \frac{b_1}{2} (J_{1,0,1}f_2 \nabla^*\rho + J_{1,1,1}f_1 \nabla^* f_1 + J_{1,2,1}\rho \nabla^* f_2 \nonumber \\
&+& J_{1,-2,-1} f_2 \nabla f^*_2 + J_{1,-1,-1} f_1 \nabla f^*_1 + J_{1,0,-1}\rho \nabla \rho \nonumber  \\
&+& J_{1,1,-1}f^*_1 \nabla f_1 
+ J_{1,2,-1} f^*_2 \nabla f_2 ) \nonumber  \\
&+&\frac{C_p}{2} e^{2i \theta_p}f^*_1, \label{eqf1} \\
\partial_t f_2 &=& \mu_2[\rho] f_2 - \frac{v_0}{2} (\nabla f_1 + \nabla^* f_3 ) \nonumber \\
&+& b_0 (\frac{C_{2,1,0}}{2}f^2_1 + C_{2,3,0} f_3 f^*_1 + C_{2,4,0} f^*_2 f_4) \nonumber \\
&+& \frac{b_1}{2} (J_{2,1,1}f_2 \nabla^* f_1 + J_{2,2,1}f_1 \nabla^* f_2 \nonumber \\
&+& J_{2,3,1}\rho \nabla^* f_3 + J_{2,-1,-1}f_2 \nabla f^*_1 + J_{2,0,-1}f_1 \nabla \rho \nonumber \\
&+& J_{2,1,-1}\rho \nabla f_1 + J_{2,2,-1} f^*_1 \nabla f_2)+C_p e^{2i \theta_p}\rho. \label{eqf2}
\end{eqnarray}
Since $\mu_3 < 0$ and $\mu_4<0$ hold for the most cases, we assume $f_3$ and $f_4$ are slaved to
$f_1$ and $f_2$ and written as
\begin{eqnarray}
f_3 &=& [(v_0 -\rho_0 b_1 J_{3,2,-1}) \nabla f_2 -2 b_0 C_{3,2,0} f_1 f_2  \nonumber \\
&-&J_{3,0,-1}f_2\nabla \rho - 3C_p e^{2i \theta_p} f_1]/(2\mu_3[\rho_0]), \label{eqf3} \\
f_4 &=& -\frac{b_0 J_{4,2,0}}{\mu_4[\rho_0]} f^2_2 - \frac{2C_p}{\mu_4[\rho_0]} e^{2i \theta_p} f_2 .  \label{eqf4}
\end{eqnarray}
Here the coefficients are defined as
\begin{eqnarray}
C_{k,q,n} &=& J_{k,q,n} + J_{k,k-q,n} \\
\mu_k[\rho] &=& P_k - 1 + a[(-1)^k - 1] + \rho b_0 C_{k,k,0}
\end{eqnarray}
Evaluating the collision integral and substituting Eq.(\ref{eqf3}) and (\ref{eqf4}) into Eq.(\ref{eqf1}) and (\ref{eqf2})  we obtain
\begin{eqnarray}
\partial_{t} f_0 &=& -\frac{1}{2} v_0 (\nabla^* f_1 +\nabla f_1^*), \\
\partial_{t} f_1 &=& (-\alpha[\rho] - \beta |f_2|^2  )f_1  \nonumber \\
&+& \sigma f_1^* f_2 - \pi_0[\rho] \nabla \rho  - \zeta[\rho] \nabla^* f_2 \nonumber \\
&+& \gamma_2 f_2 \nabla f_2^* - \lambda_n f_2 \nabla^* \rho + \lambda_1 f_1 \nabla^* f_1 + \lambda_2 f_1 \nabla f_1^* \nonumber\\
&+& \lambda_3 f_1^* \nabla f_1 
+ \gamma_1 f_2^* \nabla f_2 + \frac{C_p}{2} e^{2i \theta_p}f^*_1 ,  \\
\partial_{t} f_2 &=& (\mu[\rho] + \tau |f_1|^2 -\xi |f_2|^2) f_2 + \omega f_1^2 + \nu \Delta f_2 \nonumber \\
&-& \pi_1[\rho] \nabla f_1 +  \chi_1 \nabla^* (f_1 f_2) + \chi_2 f_2 \nabla^* f_1 + \chi_3 f_2 \nabla f^*_1 \nonumber \\
&+& \kappa_1 f^*_1 \nabla f_2 + \kappa_2 f_1 \nabla \rho + C_p e^{2i \theta_p}\rho.
\end{eqnarray}

\subsection{Coefficients of hydrodynamic equations}

The coefficients for $f_1$ equations were calculated. We set $b_0=1$.
\begin{eqnarray}
\alpha[\rho] &=& \rho \frac{4}{\pi}(\frac{4}{3} - P_1) + 1 - P_1 +2a \nonumber \\
\beta &=& \frac{8}{3\pi}(P_1 -\frac{2}{7})\frac{\frac{4}{\pi}(P_3 + \frac{4}{5})}{\rho_0 \frac{272}{35\pi}-P_3 + 1 +2a} \nonumber \\
\sigma &=& \frac{16}{5\pi}  \nonumber \\
\pi_0[\rho] &=& \frac{v_0}{2} + \frac{b_1}{4}(\pi - 2 \sqrt{2} P_1) \rho > 0, \nonumber \\
\zeta[\rho] &=& \frac{v_0}{2} - \frac{b_1}{6}\sqrt{2} P_1 \rho, \nonumber \\
\gamma_1 &=& \frac{4}{3\pi}(P_1 -\frac{2}{7})\frac{v_0 - \sqrt{2}b_1 P_3 \rho_0}{\rho_0 \frac{272}{35\pi}-P_3 + 1 +2a} + \frac{b_1}{6}\sqrt{2} P_1,  \nonumber \\
\gamma_2 &=& -\frac{b_1}{10}\sqrt{2} P_1 <0,  \nonumber  \\ 
\lambda_n &=& \lambda_1 = \lambda_2 = \frac{b_1}{4}(\pi - 2\sqrt{2} P_1)>0,  \nonumber  \\
\lambda_3 &=& -\frac{b_1}{6}\sqrt{2}P_1 <0. \nonumber 
\end{eqnarray}
where $b_1$ is related to the torque due to pairwise repulsive interaction defined as
$b_m = \int_0^{\infty} s^{m+1} g(s) ds$ .
The coefficients in the $f_2$ equation were calculated to be:
\begin{eqnarray}
\mu[\rho] &=& \rho\frac{16}{3\pi} \bigl(P_2 (2\sqrt{2} - 1) -\frac{7}{5} \bigr) + P_2 - 1 \nonumber \\
\xi &=& \frac{8}{\pi}(P_4 + \frac{1}{15}) \frac{\frac{16}{35\pi} \bigl (\frac{13}{9} -P_2(1+6\sqrt{2}) \bigr) }{-\frac{16}{15\pi}(P_4 +\frac{155}{21})\rho_0 +P_4 - 1}  \nonumber \\
\tau &=& \frac{16}{15\pi} \bigl(\frac{19}{7} - P_2(1 + \sqrt{2}) \bigr) \frac{\frac{4}{\pi} (P_3 +\frac{4}{5})}{\rho_0 \frac{272}{35\pi} - P_3 + 1 + 2a}  \nonumber \\
\omega &=&  \frac{8}{\pi}(\frac{1}{3} - P_2(\sqrt{2} - 1))  \nonumber \\
\nu &=& \frac{1}{4 (\rho_0 \frac{272}{35\pi} - P_3 + 1 + 2a)}(v_0 +b_1 P_2 \rho_0)(v_0 - \sqrt{2}b_1P_3 \rho_0) \nonumber \\
\pi_1[\rho] &=& \frac{v_0}{2} + \frac{b_1}{2} P_2 \rho >0,\nonumber 
\end{eqnarray}
\begin{eqnarray}
\kappa_1 &=& \frac{b_1}{2}P_2 - \frac{8}{15\pi} \bigl (\frac{19}{7} - P_2(1 + \sqrt{2}) \bigr) \frac{v_0 - \sqrt{2}b_1P_3 \rho_0}{\rho_0 \frac{272}{35\pi}\rho_0 - P_3 + 1 + 2a}  \nonumber \\
\kappa_2 &=& -\frac{b_1}{4}(\pi - 2P_2) \nonumber \\
\chi_1 &=& - \frac{2}{\pi} (P_3 +\frac{4}{5})\frac{v_0 +b_1 P_2 \rho_0}{\frac{272}{35\pi}\rho_0 - P_3 + 1 + 2a}  + \frac{b_1}{2}\rho_0 P_1 \nonumber \\
\chi_2 &=& \frac{b_1}{4}(\pi - 4 P_2) \nonumber \\
\chi_3 &=&  \frac{b_1}{4}(\pi - 2P_2) \nonumber
\end{eqnarray}

\subsection{Vectorial form of continuous equations}

It is useful to rewrite the full PDEs in a vector form
with using a notation for the Fourier components as
\begin{equation}
 \bm{w} \equiv \rho \bm{u}/v_0 =
\left(
\begin{array}{r}
\mathcal{R}f_1 \\
\mathcal{I}f_1   
\end{array}
\right), \quad
 \tilde{\bm{Q}} \equiv \rho \bm{Q} =
\left(
\begin{array}{rr}
\mathcal{R}f_2 &  \mathcal{I}f_2 \\
\mathcal{I}f_2  & - \mathcal{R}f_2  
\end{array}
\right). \nonumber
\end{equation}
Then, the full PDEs read
\begin{eqnarray}
\partial_t \rho &=& - v_0 \nabla \cdot \bm{w} , \\
\partial_t \bm{w} &-& 2\lambda \bm{w} (\nabla \cdot \bm{w}) - \lambda_3 (\bm{w}\cdot \nabla)\bm{w} - \lambda_3 (\bm{w} \times \nabla) \times \bm{w}   \nonumber \\
&=&  \bigl( -\alpha - \beta \tilde{S}^2 + \sigma \bm{\tilde{Q} + \frac{C_p}{2}\bm{Q}^p} \bigr) \bm{w} +  (- \zeta + \gamma_2 \bm{\tilde{Q}}) \nabla \cdot \bm{\tilde{Q}} \nonumber \\
&+& \gamma_1 [(\bm{\tilde{Q}} \cdot \nabla) \bm{\tilde{Q}}]^T - (\pi_0 + \lambda_n \bm{\tilde{Q}})\nabla \rho, \label{eqwS} \\
\partial_t \bm{\tilde{Q}} &=& \bigl(\mu +\tau |\bm{w}|^2- \xi  \tilde{S}^2\bigr)\bm{\tilde{Q}} + 
\omega \bm{\Phi} + \nu \nabla^2 \bm{\tilde{Q}}+ \chi_1 (\bm{w}\cdot \nabla) \bm{\tilde{Q}} \nonumber \\
&-& 2 \pi_1 \bm{E} + (\chi_3 -\chi_2-\chi_1)( \bm{\Omega} \cdot \bm{\tilde{Q}} - \bm{\tilde{Q}} \cdot  \bm{\Omega}) + C_p \bm{Q}^p \rho  \nonumber \\
&+&  (\chi_1+\chi_2+\chi_3)\bm{\tilde{Q}}(\nabla \cdot \bm{w}) + \chi_1 \bm{\mho} \cdot \bm{\tilde{Q}} \nonumber \\
&+& \kappa_1 \bm{\Lambda} \cdot \bm{\tilde{Q}} + \kappa_2 \bm{V} \cdot \bm{\Pi}, \label{eqQS}
\end{eqnarray}
here $\Phi_{ij} = 2 w_i w_j - \delta_{ij}|\bm{w}|^2$, $\bm{E}$ is a tensor defined by $E_{ij} = \tfrac{1}{2}(\partial_i w_j + \partial_j w_i - \delta_{ij}\nabla \cdot\bm{w})$ which coincides with the symmetric
strain rate tensor when the incompressible condition ($\nabla \cdot \bm{w}=0$) is satisfied.
and $\bm{\Omega}$ is the anti-symmetric rate of rotation tensor defined by $\Omega_{ij} = \tfrac{1}{2}(\partial_i w_j - \partial_j w_i)$, and
 $\mho_{ij} = w_j\partial_i - w_i \partial_j$, 
 $\Lambda_{ij} = \mho_{ij} + \delta_{ij}(\bm{w} \cdot \nabla)$, 
and 
\begin{equation}
\bm{V} =
\begin{pmatrix} w_x & -w_y \\ w_y & w_x \end{pmatrix} , \quad
\bm{\Pi} =
\begin{pmatrix} \partial_x \rho & \partial_y \rho \\ \partial_y \rho & -\partial_x \rho \end{pmatrix}.
\end{equation}

As explained in the main text, we neglect terms containing $\bm{w}$ twice, cancel the $f_2\triangledown^*\rho$ term to zero,
and set $\kappa_1=\kappa_2=0$, obtaining the simplified PDEs
\begin{eqnarray}
\partial_t \rho &=& - v_0 \nabla \cdot \bm{w} , \\
\partial_t \bm{w} &=& \bigl(-\alpha - \beta \tilde{S}^2 +\sigma \bm{\tilde{Q}} + \frac{C_p}{2} \bm{Q}^p\bigr) \bm{w} - \zeta \nabla \bm{\tilde{Q}}   \\
&+& \gamma_2 \bm{\tilde{Q}} (\nabla \cdot \bm{\tilde{Q}}) + \gamma_1 [(\bm{\tilde{Q}} \cdot  \nabla) \bm{\tilde{Q}}]^T - (\pi_0 +\lambda_n \bm{\tilde{Q}})\nabla \rho, \nonumber
\end{eqnarray}
\begin{eqnarray}
\partial_t \bm{\tilde{Q}} &=& \bigl(\mu - \xi  \tilde{S}^2\bigr)\bm{\tilde{Q}} + \nu \nabla^2 \bm{\tilde{Q}}+ \chi_1 (\bm{w}\cdot \nabla) \bm{\tilde{Q}} -2 \pi_1 \bm{E} \nonumber \\
&+& (\chi_3 -\chi_2-\chi_1)( \bm{\Omega} \cdot \bm{\tilde{Q}} - \bm{\tilde{Q}} \cdot  \bm{\Omega}) + C_p \bm{Q}^p \rho  \nonumber \\
&+&  (\chi_1+\chi_2+\chi_3)\bm{\tilde{Q}}(\nabla \cdot \bm{w}) + \chi_1 \bm{\mho} \cdot \bm{\tilde{Q}}.
\end{eqnarray}
When the incompressibility condition holds, the final form is similar to the Beris-Edwards equation for liquid crystals except for the linear and nonlinear active forces.
This comparison tells us that the full PDEs derived by the Boltzmann-type approach provide us with some intricate terms, including the case for larger polarity (larger net velocity) and subtle effects related to compressibility.

\subsection{Effect of nonlinear active forces (fixed $\tilde{\bm{Q}}$)}

In the steady state with given $\boldsymbol{Q}$ field of +1 topological defects, Eq. (13) in polar coordinates can be written in radial and azimuthal components, 
\begin{eqnarray}
f^a_r &=& [(\rho S(r))'+\frac{2\rho S(r)}{r}][-\zeta \cos2\theta_0 + \gamma_2 \rho S(r)] \nonumber \\
&+& [(\rho S(r))'-\frac{2\rho S(r)}{r}]\gamma_1 \rho S(r).\label{eq3} \\
f^a_\phi &=& -[(\rho S(r))'+\frac{2\rho S(r)}{r}]\zeta \sin2\theta_0.\label{eq4}
\end{eqnarray}

Except for the very vicinity of the center of the topological defect,
one can neglect $S'(r)$ and look at the force profile outside the core region. In such a case, the radial force term leads to
\begin{equation}
    f_r(r) \sim  \bigl[ \pm \zeta - (\gamma_1 - \gamma_2)\rho S \bigr] \frac{2\rho S}{r},
\end{equation}
where $+$ sign is for the target and $-$ sign is for the aster.

Based on the result in Sec.B, $\gamma_2 < 0$ holds for the systems with repulsive interactions, and $\gamma_1 > 0$ generally holds in most cases. Therefore, these two nonlinear active forces point towards the centers of defects for positive integer defects, which contributes to inward flows and accumulation.

\subsection{Conditions for accumulation and depletion}

Since we consider the case for the spontaneous formation of nematic order ($\mu>0$), 
polarity field should not form spontaneously ($\alpha>0$).
Therefore, we can assume $\alpha + \beta \tilde{S}^2>0$ since $\alpha$ and $\beta$ are positive (SI, II.B). Under the weak velocity approximation condition,  
$\tilde{\bm Q} = \rho S {\bm Q}^p$ and $\rho S \sim \sqrt{\mu/\xi} + \frac{\rho C_p}{2\mu} +\mathcal{O}(C_p^2)$, the anisotropic friction coefficient in Eq.(14) can be defined as,
\begin{equation}
    \Gamma(\tilde{\bm{Q}}) = \gamma_0 (\bm{I} - \varepsilon \tilde{\bm{Q}}),
\end{equation}
where $\gamma_0 \equiv \alpha + \beta \tilde{S}^2$ and $\varepsilon =[\sigma + C_p/(2 \rho S)] /\gamma_0$. 
The inverse of $\Gamma$ leads to
\begin{equation}
\Gamma(\tilde{\bm{Q}})^{-1} = \frac{1}{\gamma_0 (1 - \varepsilon^2 \tilde{S}^2)}(\bm{I} + \varepsilon \tilde{\bm{Q}}).
\end{equation}
Then, when $\bm{w}$ is in the steady state ($\partial_t \bm{w} =0$), $\bm{w}$ can be solved as
\begin{equation}
\bm{w} = \Gamma(\tilde{\bm{Q}})^{-1} [\bm{f}^a - \pi_0 \nabla \rho ]. \label{eqwr}
\end{equation}
Therefore, as far as $\varepsilon$ satisfies $0<\varepsilon \tilde{S} < 1$, 
anisotropic friction will not change the sign of the radial component of $\bm{w}$ for target and aster patterns and most of the spiral patterns.
If cells perfectly align to the aster (target) pattern,
anisotropic mobility, $\Gamma^{-1}(\bm{Q})$, becomes scalar: $(1+\varepsilon)/\Delta$ for the aster and $(1-\varepsilon)/\Delta$ for the target, with $\Delta = \gamma_0(1-\varepsilon^2 \tilde{S}^2)$.

Taking the inner product with $\bm{e}_r$, no flux condition is given by $f_r - \pi_0 d\rho/dr =0$.
This condition $f_r = \pi_0 d\rho/dr$ assures a steady state in the mass conservation equation in Eq.(9).
Accumulation ($\rho'<0$) or depletion ($\rho'>0$) is thus determined by the sign of $f_r$.
 
\section{Relation to other theories}

\renewcommand{\figurename}{Fig.}
\renewcommand{\thefigure}{S1}
\begin{figure}[b!]
        \centering
        \includegraphics[keepaspectratio, width=7cm]{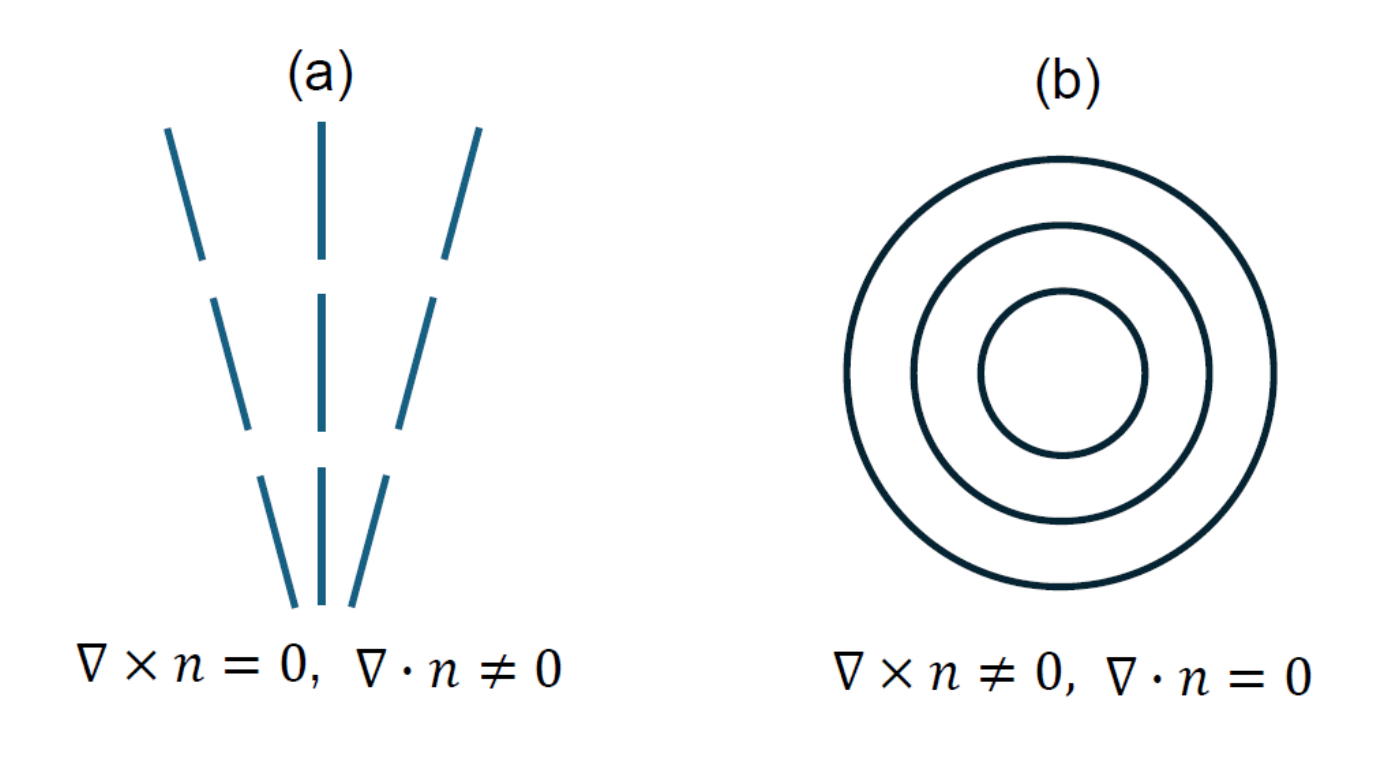}
        \caption{(a): Pure splay mode. (b): Pure bend mode}
        \label{bend_and_splay}
\end{figure}

\renewcommand{\figurename}{Fig.}
\renewcommand{\thefigure}{S2}
\begin{figure*}[t!]
        \centering
        \includegraphics[keepaspectratio,width=\textwidth]{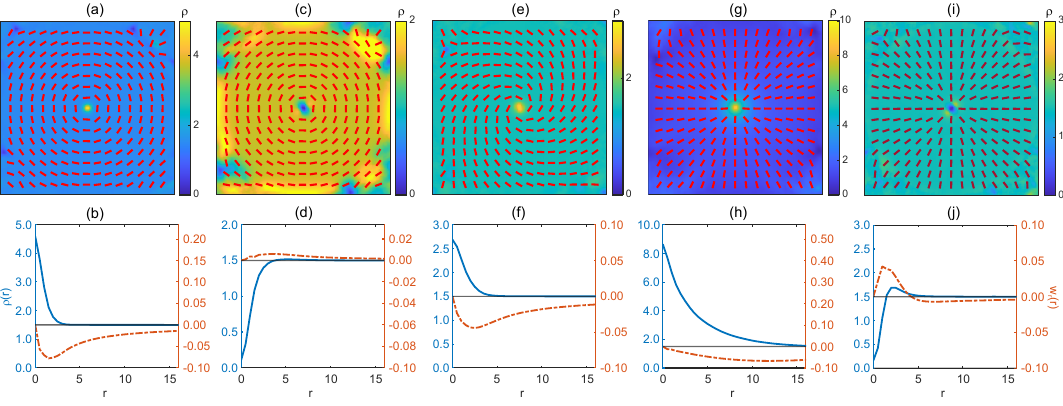}
        \caption{
        Typical snapshots of density and orientation fields (top row) and azimuthally-averaged density and velocity $w_r(r)$ profiles (bottom row)
        in a steady state of simulations of the simplified PDEs.
        (a,b): Target pattern with accumulation at the core ($v_0=0.1, b_1=0.5$).
        (c,d): Target pattern with depletion at the core ($v_0=0.4, b_1=0.2$). 
        (e,f): Spiral pattern ($\theta_0=\pi/3$) with accumulation at the core ($v_0=0.1, b_1=0.4$).
        (g,h): Aster pattern with accumulation at the core ($v_0=0.5, b_1=0.25$).
        (i,j): Aster pattern with depletion at the core ($v_0=0.05, b_1=0.6$).
        For all cases above, $\rho_0=1.5, a=0.4, \eta=0.2, \nu=0.2, L=64$. The net velocity is weighted by $\rho$.
        }
        \label{fig_S2}
\end{figure*}

\subsection{Relation to the director representations}

We note that one of the nonlinear terms, $\bm{Q}(\nabla \cdot \bm{Q})$, was discussed in the context of active suspension under strong constraint in 2D as a stabilizing active force\cite{Maitra2018}.
It is also helpful to discuss the relationship with the nematic contributions in polar systems considered in earlier works~\cite{Blanch-Mercader2021E}.
Although in our approach, $\tilde{\bm{Q}}$ is density-weighted and contains the amplitude $S$,
if we assume $S=1$ and uniform density, it is possible to rewrite as $Q_{ij}\simeq 2(n_i n_j - \tfrac{1}{2}\delta_{ij})$ with the director field $\bm{n}$. In such an approximation, $\tilde{\bm{Q}} (\nabla \cdot \tilde{\bm{Q}})$ can be written with the director field as.
\begin{eqnarray}
    Q_{ij}\partial_k Q_{jk} &\simeq& 4(n_i n_j - \frac{1}{2}\delta_{ij})\partial_k (n_j n_k -  \frac{1}{2}\delta_{jk}) \nonumber \\
    &=& 4 n_i n_j n_j \partial_k n_k + 4 n_i n_k \partial_k (\frac{1}{2}n_j^2) -2 n_i \partial_k n_k  \nonumber \\
    &-& 2n_k \partial_k n_i = 2 (n_i \partial_k n_k - n_k \partial_k n_i ),
\end{eqnarray}
where the relation, $n_j n_j = 1$ and $\partial_k (n_j^2)=0$, are used.
Similarly, $(\tilde{\bm{Q}} \cdot \nabla)\tilde{\bm{Q}}$ term can be written as, 
\begin{eqnarray}
    Q_{jk}\partial_k Q_{ij} &\simeq& 4(n_j n_k -  \frac{1}{2}\delta_{jk})\partial_k (n_i n_j -  \frac{1}{2}\delta_{ij}) \nonumber \\
    &=& 4 n_j n_k \partial_k (n_i n_j) - 2 \delta_{jk} \partial_k (n_i n_j)  \nonumber \\
    &=& 2 (n_k \partial_k n_i -n_i \partial_k n_k).
\end{eqnarray}
Therefore,
\begin{equation}
    \gamma_1 Q_{jk}\partial_k Q_{ij} + \gamma_2 Q_{ij}\partial_k Q_{jk} \simeq
    2(\gamma_1 - \gamma_2) (n_k \partial_k n_i -n_i \partial_k n_k).
\end{equation}
Under this approximation, 
 the linear active force term, $-\zeta \nabla \cdot \tilde{\bm{Q}}$ is also written as
\begin{equation}
-2\zeta \partial_k (n_i n_k - \frac{1}{2}\delta_{ik}) = -2\zeta (n_k \partial_k n_i +n_i \partial_k n_k).
\end{equation}
Thus, the sum of active force terms can be expanded by the two contributions as
\begin{eqnarray}
    -\zeta &\nabla_k& Q_{ik}  + \gamma_1 Q_{jk} \partial_k Q_{ij} + \gamma_2 Q_{ij}\partial_k Q_{jk} \nonumber \\
   &\simeq& A ( n_k \partial_k n_i +n_i \partial_k n_k) + B (n_k \partial_k n_i -n_i \partial_k n_k) \nonumber \\
    &=& (A+B) n_k \partial_k n_i + (A-B) n_i \partial_k n_k ,  \label{eqbendsplay}
\end{eqnarray}
where $A = -2\zeta$ is the coefficient of symmetric force terms and $B = 2(\gamma_1 - \gamma_2)$
is the coefficient of the anti-symmetric force terms.

In our model, $A \pm B = 0$ does not hold except at the transition lines; therefore, the two terms in the third line of Eq.(\ref{eqbendsplay}) generally exist and can be viewed as an expansion by pure bend and pure splay modes, by using the relation $n_k \partial_k n_i = -[\bm{n}\times (\nabla \times \bm{n})]_{i}$ in 2D for bend,
and $n_i \partial_k n_k = [\bm{n} (\nabla \cdot \bm{n})]_i$ for splay.
Note that $\nabla \cdot \bm{n}=0$ holds for pure bend, $\bm{n} = (-\sin \theta, \cos \theta)$, and $\nabla \times \bm{n} =0$ holds for pure splay mode, $\bm{n} = (\cos \theta, \sin \theta)$.
(See Fig.S1)

The second line in Eq.(\ref{eqbendsplay}) represents an expansion by the symmetric and anti-symmetric terms consisting of the second order in $\bm{n}$ and the first order in the derivative.
Interestingly, the usual active nematic theories neglected the anti-symmetric term proportional to the $\gamma_1 -\gamma_2$.

The expansion of the active forces by bend and splay modes in 2D provides an intuitive description. Suppose the expansion coefficients of $n_k \partial_k n_i$ and $n_i \partial_k n_k$ are equal, as in the one constant approximation often used in liquid crystal theories. In that case, there is no asymmetric term in the second line of Eq.(\ref{eqbendsplay}). However, an anti-symmetric ($\gamma_1 -\gamma_2$) term generally exists.

\subsection{Density profile and velocity around +1 defects}

Figure S2 shows simulation results of the simplified PDEs for target, aster, and spiral defects.